%% file: NSresponseNewton.tex
\documentclass[
	aps, prd, reprint,
	10pt, notitlepage, a4paper,
        floats, floatfix,
	amsmath, amssymb, amsfonts, eqsecnum,
	superscriptaddress,
	showpacs, showkeys,
	nofootinbib,
]{revtex4-1}

\usepackage{ifthen}
\newboolean{prd}
\setboolean{prd}{false}
\newboolean{arxiv}
\setboolean{arxiv}{true}

\newboolean{notprd}
\setboolean{notprd}{true}
\ifprd
\setboolean{notprd}{false}
\fi
\newboolean{notarxiv}
\setboolean{notarxiv}{true}
\ifarxiv
\setboolean{notarxiv}{false}
\fi

\usepackage{graphics,graphicx}
\usepackage[caption=false]{subfig}
\usepackage[usenames]{xcolor}

\ifnotprd
\xdefinecolor{mylinkcolor}{rgb}{0,0,0.5}
\usepackage[
	bookmarksnumbered, bookmarksopen, bookmarksopenlevel=2,
\ifnotarxiv
	breaklinks=true,
\fi
	colorlinks=true, filecolor=mylinkcolor, citecolor=mylinkcolor,
	linkcolor=mylinkcolor, urlcolor=mylinkcolor, menucolor=mylinkcolor,
]{hyperref}
\else

\fi

\usepackage{wasysym}

\newcommand{\be}{\begin{equation}}
\newcommand{\ee}{\end{equation}}
\newcommand{\bea}{\begin{eqnarray}}
\newcommand{\eea}{\end{eqnarray}}

\def\vct#1{\mathbf{#1}}
\def\vctsym#1{\boldsymbol{#1}}

\def\nl{\\ & \quad}

\DeclareMathOperator{\Order}{\mathcal{O}}

\allowdisplaybreaks

\begin{document}

% for PRD one has to get rid of hyperref
\ifnotprd
\hypersetup{
	pdftitle={Effective action and linear response of compact objects in Newtonian gravity},
	pdfauthor={Sayan Chakrabarti, Terence Delsate, Jan Steinhoff}
}
\fi

\title{Effective action and linear response of compact objects in Newtonian gravity}

\author{Sayan Chakrabarti}
% \email{sayan.chakrabarti@ist.utl.pt}
\affiliation{Centro Multidisciplinar de Astrof\'isica --- CENTRA, Departamento de F\'isica,
	Instituto Superior T\'ecnico --- IST, Universidade de Lisboa --- ULisboa,
	Avenida Rovisco Pais 1, 1049-001 Lisboa, Portugal, EU}
\affiliation{Department of Physics, Indian Institute of Technology Guwahati, Assam 781039, India}

\author{T\'erence Delsate}
% \email{terence.delsate@umons.ac.be}
\affiliation{Centro Multidisciplinar de Astrof\'isica --- CENTRA, Departamento de F\'isica,
	Instituto Superior T\'ecnico --- IST, Universidade de Lisboa --- ULisboa,
	Avenida Rovisco Pais 1, 1049-001 Lisboa, Portugal, EU}
\affiliation{UMons, Universit\'e de Mons, Place du Parc 20, 7000 Mons, Belgium, EU}

\author{Jan Steinhoff}
\email[Corresponding author: ]{jan.steinhoff@ist.utl.pt}
% \thanks{primary investigator}
% \homepage{http://jan-steinhoff.de/physics/}
\affiliation{Centro Multidisciplinar de Astrof\'isica --- CENTRA, Departamento de F\'isica,
	Instituto Superior T\'ecnico --- IST, Universidade de Lisboa --- ULisboa,
	Avenida Rovisco Pais 1, 1049-001 Lisboa, Portugal, EU}
\affiliation{ZARM, University of Bremen, Am Fallturm, 28359 Bremen, Germany, EU}

\date{\today}

\begin{abstract}
We apply an effective field theory method for the gravitational interaction of compact
stars, developed within the context of general relativity, to Newtonian gravity.
In this effective theory a compact object is represented by a point particle
possessing generic gravitational multipole moments. The time evolution of the
multipoles depends on excitations due to external fields. This can formally be
described by a response function of the multipoles to applied fields. The poles
of this response correspond to the normal oscillation modes of the star. This
gives rise to resonances between modes and tidal forces in binary systems. The
connection to the standard formalism for tidal interactions and resonances in
Newtonian gravity is worked out. Our approach can be applied to more complicated
situations. In particular, a generalization to general relativity is possible.
\end{abstract}


\maketitle

\section{Introduction}
The Newtonian description of gravitating systems was formulated in the 17th
century and is nowadays part of most classical mechanics textbooks.
Remarkably enough, it is still possible to
understand new facets of this model. At the same time, Newton's formulation of
gravity is mathematically simple and allows straightforward analytic treatment.
% In particular, the Poisson equation gives very simple solutions, even when including
% nontrivial angular dependence.
In particular, the Poisson equation admits a well understood Green function solution.
This is why the Newtonian
theory is also a good starting point to investigate some fundamental
aspects of general relativity.

One of these aspects, among many others, is tidal interaction in a binary system.
In the beginning of the last century, A.~E.~H.~Love introduced
two numbers named after him in order
to characterize the shape change of the Earth due to an external tidal
potential \cite{Love:1911}. A third Love number was introduced later by Shida \cite{Shida:1912}.
These numbers essentially encode the mass redistribution of a planet due to
tidal forces, including those generated by the tidal bulges themselves.
A tide-generating potential arises, for example, in binary systems, such as binary stars
or a planet and its satellite, e.g., the Earth and the moon.
The latter example comprises ocean tides, which motivated the first studies of
the subject and its naming. Tidal forces also play a crucial role for
the concept of local inertial frames in general relativity. Despite this
close interrelation between tidal interaction and general relativity, a relativistic
definition of Love numbers was worked out only a century after its Newtonian
counterpart \cite{Hinderer:2007, Damour:Nagar:2009:4, Binnington:Poisson:2009}; see also
Refs.\ \cite{Flanagan:Hinderer:2007, Berti:Iyer:Will:2007}. It should be emphasized that
modeling tidal interactions through constant Love numbers assumes a slowly
(adiabatically) varying tidal field.

On another hand, binary systems possess a typical frequency due to the revolution of the bodies around
each other. In some situations, particularly for binary stars, the orbital frequency can
become comparable to the normal oscillation frequencies of one of the binary's components.
Therefore, a resonance can take place, possibly leading to large energy and angular momentum transfers.
The analytic framework to treat dynamical (time-dependent) tidal interactions in
Newtonian gravity goes back to Press and Teukolsky \cite{Press:Teukolsky:1977}. Though they focus on tidal capture
as an application, resonances in bound binaries can be treated as well. Tidal
heating, tidal disruption,
and tidal locking are other astrophysically
important effects
belonging into this domain, see, e.g., Refs.\ \cite{Meszaros:Rees:1992,
Lai:1994, Gingold:Monaghan:1980, Bildsten:Cutler:1992}.
A more complete list of references for generic
binaries is given in, e.g., Refs.\ \cite{Alexander:1987, Flanagan:Racine:2007}.
Also, resonances in neutron star binaries have been considered using these methods
\cite{Shibata:1993, Reisenegger:Goldreich:1994, Lai:1994, Kokkotas:Schafer:1995,
Ho:Lai:1998, Flanagan:1998, Lai:Wu:2006, Flanagan:Racine:2007}, though one should be
careful to draw final conclusions from such investigations, as general relativistic corrections can
be large in this case.

Our motivation is indeed to improve the situation for neutron star binaries by formulating
a general model for time-dependent tides in general relativity.
But the problem is highly nontrivial due to the nonlinear nature of general relativity.
Here we focus on the Newtonian formulation of the problem instead, because it is
described in terms of simple functions, and most of the technical difficulties of
the relativistic case are avoided. Though excellent formalisms for dynamical tidal
interactions already exist \cite{Press:Teukolsky:1977, Rathore:Broderick:Blandford:2002,
Alexander:1987}, we devise yet another formalism based on quantities that
allow a general relativistic generalization.
Our reformulation is thus also interesting in its own right. For instance, it can
turn out to be advantageous for more complicated situations in the Newtonian regime, too, e.g.,
when nonlinear tidal perturbations and mode coupling are considered (see
Refs.\ \cite{Weinberg:Arras:Burkhart:2013, Weinberg:Arras:Quataert:Burkhart:2011}
for recent investigations on neutron stars).
The relativistic case is presented in another publication \cite{Chakrabarti:Delsate:Steinhoff:2013:2}.

Our method relies on an effective field theory approach to gravitational interaction
of compact objects in classical general relativity \cite{Goldberger:Rothstein:2006}, which
is cut down to the Newtonian case for the present work.
The principle is to effectively represent a compact source, with all its potentially
complicated internal dynamics, by a point-particle source decorated by multipolar degrees of freedom
\cite{Goldberger:Rothstein:2006, Goldberger:Rothstein:2006:2, Goldberger:Ross:2009}.
The dynamics of these multipoles can be encoded through a response function to external
gravitational fields \cite{Goldberger:Rothstein:2006:2}.
The game is then to match the gravitational field of the effective source
to the gravitational field of the actual compact object, which fixes the response.
The response function, which is in fact the retarded propagator of the
multipolar degrees of
freedom, then encodes all the (macroscopically relevant) time-dependent internal dynamics.
The advantage is that it is sufficient, at least in linear perturbation theory, to know the
gravitational field of a \emph{single perturbed} compact object in order to make
predictions about a gravitationally interacting many-body system. These predictions
can be derived from the effective theory, e.g., with the help of diagrammatic
techniques developed for quantum field theory (Feynman diagrams), see
Refs.\ \cite{Damour:EspositoFarese:1995, Goldberger:Rothstein:2006} and references therein
for the case of relativistic gravitational interaction.
% In fact, the approach is
% the same as in Quantum Mechanics where, given the response of a single particle
% to external perturbations, the dynamics of interacting particles can be inferred
% perturbatively.

The normal oscillation modes of a compact object play a crucial role in our
approach.  A compact configuration can possess a series of normal modes, and each
mode can be excited by an external field of the appropriate frequency.
Indeed, the mode frequencies maximize the response to external
perturbations, similar to a forced oscillator. 
In fact, the multipolar degrees of freedom turn out to be sums of these
forced oscillators. From this point of view, multipoles are composite, and the
fundamental composing quantities are the deformation amplitude modes (this will
be detailed later).
%In fact, in the Newtonian case, it can be viewed as a sum of forced
%oscillators, one for each mode.
All this treatment can be carried out explicitly; i.e., an analytic expression
for the response function can be given in terms of oscillation mode frequencies
and overlap integrals, which are defined in the standard approach \cite{Press:Teukolsky:1977,
Rathore:Broderick:Blandford:2002, Alexander:1987}. But the concepts of effective
action, response functions/propagators, matching, and perturbations of single compact
objects can be generalized to the general relativistic case
\cite{Chakrabarti:Delsate:Steinhoff:2013:2}, which is less obvious for some elements
of the standard formalism. As we work with an effective action here, it is most natural
to make connections to the standard approach through the variational principle in
Ref.\ \cite{Rathore:Broderick:Blandford:2002}, which was developed to describe mode resonances
in binary white dwarfs systems \cite{Rathore:Blandford:Broderick:2004, Rathore:2005}.
% Our goal here is to describe
% the whole dynamics of the central body by a point surrounded with multipolar
% degrees of freedom. This can be achieved provided the Newtonian potentials of
% the point multipoles and of the central object with all its internal dynamics
% are the same outside the compact object. Interestingly, the procedure in order
% to match these two systems involves computing the response of the central
% object to external perturbations.

% This approach is particularly useful for Neutron Star. In fact, most
% of the conclusions regarding the physical aspects are already known, but here we
% formulate the problem differently, and show how it connects directly to the
% relativistic case, and we further give insights on the understanding of
% multipoles in gravity theories.

Resonances between oscillation modes and orbital motion can also be interesting
for future gravitational wave astronomy, in particular for binaries comprising
neutron stars. Past investigations suggest that these resonances are likely not
detectable through gravitational waves using observatories currently under construction
\cite{Reisenegger:Goldreich:1994, Balachandran:Flanagan:2007}.
However, it should be stressed again that such an analysis did not model the internal
dynamics entirely within general relativity. In fact, numerical relativity simulations
suggest that resonances can even be driven into the nonlinear regime in certain cases
(eccentric orbits) and thus contribute substantially to gravitational waves
\cite{Gold:Bernuzzi:Thierfelder:Brugmann:Pretorius:2011}.
Of course, tidal interactions are important during the inspiral also away from a resonance,
and they can encode important information on the nuclear equation of state in the
gravitational wave signal, which is supported by both analytic models
\cite{Baiotti:etal:2010, Damour:Nagar:Villain:2012, Hinderer:Lackey:Lang:Read:2009}
and numerical relativity
\cite{Bejger:GondekRosinska:Gourgoulhon:Haensel:Taniguchi:etal:2004,
Shibata:Kyutoku:2010, Pannarale:Rezolla:Ohme:Read:2011, Bernuzzi:Nagar:Thierfelder:Brugmann:2012}.
Also, the merger of binary neutron stars \cite{Bauswein:Janka:2011} and
the details of the tidal disruption in mixed (neutron star --black hole) binaries
\cite{Wiggins:Lai:1999, Ishii:Shibata:Mino:2005, Ferrari:Gualtieri:Pannarale:2009, Ferrari:Gualtieri:Pannarale:2008,
Kyutoku:Shibata:Taniguchi:2010,Kyutoku:Okawa:Shibata:Taniguchi:2011} allows
crucial statements on the equation of state. Furthermore, resonances can
have other observable effects in neutron star binaries besides gravitational
waves. It was suggested in Ref.\ \cite{Tsang:etal:2012}
that oscillations powered by a resonance can be strong enough to shatter the neutron
star crust. This could explain precursor flares in short gamma ray bursts, with the
main burst produced by the merger of the binary. Similar violent processes may be
observable in the electromagnetic spectrum if instabilities develop.
Indeed, some oscillation modes---including the f-mode---of rotating
neutron stars can become unstable \cite{Gaertig:Glampedakis:Kokkotas:Zink:2011}.

In spite of these interesting prospects, an entirely satisfactory analytical tidal
interaction model incorporating a general relativistic internal dynamics is still
missing.
So far resonances could only be accounted for by investigating a point mass
orbiting the star; see Refs.\ \cite{Kojima:1987, Ruoff:Laguna:Pullin:2000, Gualtieri:Berti:Pons:Miniutti:Ferrari:2001, Pons:Berti:Gualtieri:Miniutti:Ferrari:2001}
and also Ref.\ \cite{Fang:Lovelace:2005} for the black hole case.
In between a fully relativistic and a Newtonian treatment is a post-Newtonian
approximation of the internal dynamics. A corresponding formalism was developed in
Refs.\ \cite{Damour:Soffel:Xu:1991, Damour:Soffel:Xu:1992, Damour:Soffel:Xu:1993, Damour:Soffel:Xu:1994}
and further elaborated in Refs.\ \cite{Flanagan:1998, Racine:Flanagan:2005}.
The first post-Newtonian approximation was applied to binary neutron stars in Refs.\ \cite{Shibata:Taniguchi:Nakamura:1998, Taniguchi:Asada:Shibata:1998, Taniguchi:Shibata:1998}.
Another good dynamical model was developed around the Newtonian limit in
Refs.\ \cite{Maselli:Gualtieri:Pannarale:Ferrari:2012, Ferrari:Gualtieri:Maselli:2011}.
% However, it is not clear if it can account for
% resonances of dynamical tidal fields with oscillation modes of the star.
Still, it would be optimal to find the description of tidal interaction based on
quantities
derived from perturbation theory around nonlinear background solutions. So far this succeeded
only for the definition of relativistic Love numbers \cite{Hinderer:2007, Damour:Nagar:2009:4,
Binnington:Poisson:2009, Damour:Nagar:2009}, which represent the adiabatic limit of the
tidal response. (Tidal coefficients beyond the adiabatic case were formally
introduced in Refs.\ \cite{Bini:Damour:Faye:2012}.)
% Here we show how to read off the overlap integrals from the
% external field of a perturbed Newtonian star in the present paper, which generalizes to the
% relativistic case and thus defines relativistic overlap "integrals"
% \cite{Chakrabarti:Delsate:Steinhoff:2013:2}.
Further, relativistic perturbation theory was used to study absorptive effects of black holes
in tidal fields; see Refs.\ \cite{Poisson:2004, Poisson:2005, Taylor:Poisson:2008, Poisson:2009,
Comeau:Poisson:2009, Chatziioannou:Poisson:Yunes:2012} and references therein.

In this paper, we start in Sec.\ \ref{effaction} by reviewing and motivating the
effective field theory approach in general relativity and argue that the same formalism
applies in the
Newtonian theory. We explain in detail how the tidal interaction
can be encoded using time-dependent multipolar degrees of freedom and work out the
connection to tidal coefficients and absorption coefficients. Section
\ref{stellar_perturbation} reviews a variational principle for perturbations of a gravitating system in
the Newtonian theory. We improve on previous presentations by performing all
transformations at the level of the action. Section \ref{devlpmnt} discusses some useful tools, such as the symmetric trace-free (STF) tensor formalism, which is 
useful for the present work.  Next, we show how to cast stellar perturbations in
an effective theory formulation in Sec.\ \ref{micro} and show how to read
off the overlap integrals from the
external field of a perturbed Newtonian star (which
generalizes to the
relativistic case and thus defines relativistic overlap ``integrals''
\cite{Chakrabarti:Delsate:Steinhoff:2013:2}). Subsequently, we extract
the
full time-dependent dynamics of the quadrupolar degrees of freedom in Sec.\ \ref{match}
by analyzing the perturbation field of a single compact object.
We adapt the matching to a numerical setup in Sec.\ \ref{numeric} and illustrate
our results with a simple (analytic) background solution corresponding
to a particular polytropic star.
Section \ref{conclusions} discusses the conclusions and outlook.
Finally, the Appendix provides more details on some mathematical aspects of our
results. %as well as the generalization to arbitrary multipoles.

Our conventions are the following. Greek indices refer to 4-dimensional
spacetime, while lowercase Latin indices denote
spatial components. (But sometimes $l$ and $m$ are used for angular momentum
quantum numbers, which should be obvious from the context.) Uppercase Latin
indices are multi-indices of spatial components. 
Einstein's summation convention is applied to these types of indices.
Our sign convention for the metric $g_{\mu\nu}$ is taken to be $(-,+,+,+)$
and the Riemann tensor is
\begin{equation}
R^{\mu}{}_{\nu\delta\sigma} = \Gamma^{\mu}{}_{\nu\sigma,\delta}
        - \Gamma^{\mu}{}_{\nu\delta,\sigma}
        + \Gamma^{\mu}{}_{\lambda\delta} \Gamma^{\lambda}{}_{\nu\sigma}
        - \Gamma^{\mu}{}_{\lambda\sigma} \Gamma^{\lambda}{}_{\nu\delta} ,
\end{equation}
where $\Gamma^{\mu}{}_{\nu\sigma}$ is the Christoffel symbol.
We use units such that the speed of light is $c=1$.
The Newton constant is denoted $G$.
% The Newtonian potential $\Phi$ is given by $g_{00} \approx - 1 - 2 \Phi$.

\section{Effective action\label{effaction}}
The goal of this paper is to apply the ideas of effective field theory developed in
general relativity to the Newtonian theory of gravity.
In fact, we find later on that the nonrelativistic limit of the effective action
and an action derived for perturbations in pure Newtonian gravity are essentially
equivalent. This should of course be the case if the effective action is constructed
from the correct symmetries.
We first discuss various forms of effective actions for tidal interactions (and their
relations) in the relativistic case, since the discussion is essentially the same
in the nonrelativistic case.
Afterward, we determine the Newtonian limit and construct interaction potentials.

\subsection{Effective field theory in gravity}
Some of the most important sources of gravitational waves are the inspiral,
merger, and ringdown phases of compact objects in NS-NS, NS-BH, or BH-BH
binaries, where NS stands for neutron star and BH for black hole. These waves will contain important information about the 
compact object. Such binary systems exhibit a number of different
length scales, for example, the size of the compact objects, the orbital radius,
and the wavelength of the emitted radiation.    
On general grounds, attacking a two-body problem in general relativity needs numerical methods. However, there 
exist certain regimes with a clear separation of length scales, and
one can track the problem analytically using approximate methods. 
For instance, the post-Newtonian expansion or the extreme mass ratio
inspiral limits are two such cases. These two methods are efficient in a certain
regime but break down above some limits.
% In general the coalescence of the binary system can be studied using PN
% approximation. Although it provides a semi-analytic
% approach towards the evolution of the system, it is poorly convergent in strong gravity regime. For this reason, it is best suited for describing the inspiral 
% phase. 

There exists a systematic way to account for effects that arise at
different length scales, which is the effective field theory approach;
see, e.g., Refs.\ \cite{Georgi:1994, Burgess:2007, Rothstein:2003} for reviews.
These methods are often associated with quantum theory, and in fact the
first application to the nonrelativistic limit of gravity \cite{Donoghue:1995}
was focused on quantum corrections.
Nevertheless, the methods are very useful for astrophysical binaries and
their gravitational waves in classical gravity, too,
which was put forward in Refs.\ \cite{Goldberger:Rothstein:2006,
Goldberger:Rothstein:2006:3, Goldberger:2007}.
In this situation, the compact object is described by a worldline action
which includes all the possible terms consistent with the diffeomorphism
invariance of general relativity [and eventually symmetries inherited from the single
unperturbed object, like SO(3) rotational invariance].
The method is particularly helpful to systematically 
account for tidal and other finite size effects, which become important in some regime,
e.g., the late inspiral
stage of a compact binary system.
The effective action in Refs.\ \cite{Goldberger:Rothstein:2006,
Goldberger:Ross:2009} can be used to find the point particle
description of nondissipative finite size effects
and can be extended in order to include
% realistic astrophysical situations such as
dissipation \cite{Goldberger:Rothstein:2006:2}.
In the latter case, the gravitational multipoles enter the action as worldline degrees of
freedom, for which the dynamics is encoded in a propagator or response function.
This manifestly covariant extension of the point-mass action is adopted
as a model for extended objects here. % In order to explain in more detail where
%the present research
% project is aiming at, I will review some of these extensions now. 

Two approaches to effective theories can be contrasted \cite{Georgi:1994}:
A Wilsonian approach of integrating out short-scale physics from a
full theory and a continuum effective field theory approach, where
the form of the effective action is first constructed by hand (e.g.,
using symmetries) and afterward specialized by a matching to the full theory.
The latter method is usually much simpler.
However, the setting of the present work is simple enough to
follow an approach along the lines of Wilson's ideas, too.
This is elaborated in Sec.\ \ref{micro}.
An analytic matching to the full theory (variational fluid dynamics)
in the spirit of continuum effective theories is discussed in Sec.\ \ref{match}.
Next, we discuss in Sec. \ref{numeric} a matching procedure to
solutions of the full theory obtained numerically. This in an important shift in
paradigm,
as many complicated systems cannot be treated analytically, but numeric
simulations are usually possible. This includes stars with more
complicated (realistic) internal structure and nonlinear perturbations.
Not surprisingly, the numeric matching is the approach followed
for neutron stars in the general relativistic generalization of our work
\cite{Chakrabarti:Delsate:Steinhoff:2013:2}, while the black hole case still
allows analytic treatment.

The more standard method to handle different scales in classical
general relativity is the matched asymptotic expansion; see Refs.\ \cite{Blanchet:2006,
Blanchet:2009} and references therein. For a comparison of
matched asymptotic expansion and matching in effective field theory,
see Ref.\ \cite{Kol:Smolkin:2007}.
Another aspect of the effective field theory approach in
\cite{Goldberger:Rothstein:2006} is the consequent formulation
of perturbative calculations through Feynman diagrams. However,
diagrammatic approaches were used in classical gravity before,
see \cite{Damour:EspositoFarese:1995} and references therein.
Effective worldline actions as a model for finite size
effects were first discussed in the context of scalar-tensor theories
of gravity \cite{Damour:EspositoFarese:1998}, since they are potentially
more relevant in this class of alternative theories compared to the
general relativistic case.

\subsection{Action for dynamical multipoles}
In the following, we intend to illustrate the usefulness of the effective
action in Ref.\ \cite{Goldberger:Rothstein:2006:2} in the context of tidal interactions.
We simply start with the action constructed in Ref.\ \cite{Goldberger:Rothstein:2006:2}.
% This construction is redone in the much simpler Newtonian case in Sec.\ \ref{match}
% for completeness.

% The interaction term \eqref{tidalaction} is obviously not dynamic, in
% particular it is instantaneously aligned with the external field. A way to
% tackle this drawback is to introduce a dynamical
Introducing a dynamical quadrupole degree of freedom in the form of a STF
tensor $Q^{ab}$, the point-particle (PP) worldline action proposed
in Ref.\ \cite{Goldberger:Rothstein:2006:2}
reads
\begin{equation}
S_{\text{PP}}=\int d\tau  \left[ - m
-\frac{1}{2 } E_{ab}Q^{ab} + \ldots\right],
% L_{Q} = -\frac{ Q_{\mu\nu} E^{\mu\nu}}{(-u_\sigma u^{\sigma})^{3/2}}.
\label{eq:effective_action}
\end{equation}
where $m$ is a constant mass parameter, $E_{\mu\nu}$ is the electric
part of the Weyl tensor $C_{\mu\rho\nu\sigma}$,
$E_{\mu\nu} = C_{\mu\rho\nu\sigma} u^\rho u^\sigma$, and $u^{\mu}$ is the
4-velocity with respect to the proper time $\tau$.
(In vacuum, $C_{\mu\rho\nu\sigma} = R_{\mu\rho\nu\sigma}$ holds.)
$E_{\mu\nu}$ is evaluated at the position of the object. 
In this section, Latin indices
denote spatial components of the comoving frame in a local Cartesian basis.
The dots denote higher multipole corrections as well as magnetic-type
multipoles.
For the sake of simplicity, we restrict to interactions of the mass
quadrupole $Q^{ab}$ for now, which can be classified as electric type.
% The extension is straightforward
% \cite{Goldberger:Ross:2009} and is partly discussed in Appendix \ref{higher}.

% In the following, we will parametrize the worldline by the proper time $\tau$ where $u^2=-1$.
% % In the rest of this section, we review the basic principles of the Effective
% % Field Theoretical approach in GR introduced in \cite{Goldberger:Rothstein:2006:2}.
% Then the construction can be straightforwardly transferred to the nonrelativistic case,
% where $\tau \approx t$ and t is the coordinate time. 

Within the approach proposed in Ref.\ \cite{Goldberger:Rothstein:2006:2},
the dynamics of $Q^{ab}$ is introduced via its two-point function, or propagator,
which is obtained from a matching procedure in the frequency regime of interest.
%, cf.\ equations (22--23)
% This action was also used to describe absorptive effects via the dynamical
% quadrupole-like degrees of freedom $Q^{ab}_E = Q^{ba}_E$. %Here fields are
%evaluated at $\vct{R}_{*}(t)$. 
This provides a model for the quadrupole, which is required in order to complete the
description of the system. As the outcome of the matching is yet unknown,
the complete effective action including the quadrupole dynamics cannot be
given explicitly. However, if we restrict to linear tidal effects, the
pure quadrupole part of the action can always be written in terms of an
invertible Hermitian linear operator $\mathcal O^{ab}{}_{cd}$ in the form
\be
S_Q =-\frac{1}{2} \int d\tau \, Q_{ab} \mathcal O^{ab}{}_{cd} Q{}^{cd} .
\label{action_q}
\ee
The complete effective action for the compact object including the mass
quadrupole finally reads
\be
S_{\text{eff}} = S_Q + S_{\text{PP}} .
\ee

\subsection{Quadrupole response\label{int_response}}
% is clear instead is that whatever the dynamical part of the quadrupole model
% is, the source term is given by the effective action and is simply
% $E_{\mu\nu}$.
% In other words, given a model for the quadrupole, the equation would be of the form
As $Q^{ab}$ is assumed to be a dynamical variable here, its equation of
motion follows from $S_{\text{eff}}$ as
\be
\mathcal O^{ab}{}_{cd} Q^{cd} = -\frac{1}{2} E^{ab}(\tau),
\label{eq:formal_eq_quad}
\ee
where we used that $\mathcal O^{ab}{}_{cd}$ should be Hermitian.
If this is not the case, then one must complete the quadrupole model by
giving its equation of motion
(\ref{eq:formal_eq_quad}) instead of an action $S_Q$.
The formal solution to Eq.\ \eqref{eq:formal_eq_quad} is given by
\begin{equation}\label{Qformal}
Q^{ab}(\tau) = - \frac{1}{2} \int d \tau' \, F^{ab}{}_{cd}(\tau, \tau')
E^{cd}(\tau') ,
\end{equation}
where $F$ is a Green function or propagator of the operator $\mathcal O^{ab}{}_{cd}$.
One can make the ansatz
\begin{equation}
F^{ab}{}_{cd}(\tau, \tau') = F(\tau - \tau') \hat{\delta}^{ab}{}_{cd} ,
\end{equation}
where $\hat\delta^{ab}{}_{cd}$ is the STF projector of the
form
\begin{equation}
\hat{\delta}^{ab}{}_{cd} = \frac{1}{2} ( \delta^{a}{}_{c} \delta^{b}{}_{d}
	+ \delta^{a}{}_{d} \delta^{b}{}_{c} ) - \frac{1}{3} \delta^{ab}
\delta_{cd} .
\end{equation}

% The response function $F$ is obtained from a
% matching procedure and includes self-interactions of $Q^{ab}$. 
%% Notice that
% \begin{equation}
% E_{ij} \approx R_{0i0j} \approx \partial_i \partial_j \Phi .
% \end{equation}
% In order to do so, we make an ansatz for the quadrupole in the following form
% \begin{equation}
% Q^{ij}(t) = - \frac{1}{2} \int d t' \, F^{ij}{}_{kl}(t, t') E^{kl}(t') .
% \end{equation}
We require that $F^{ab}{}_{cd}(\tau, \tau') = 0$ if $\tau'>\tau$, so we consider
the retarded Green function. Then, $F^{ab}{}_{cd}(\tau, \tau')$ provides
the response function of the quadrupole under external gravitational forces.
This solution implements the boundary condition that the quadrupole vanishes in the
absence of external forces and for $\tau \rightarrow - \infty$.

It is often useful to consider the response functions in the frequency domain.
In particular, one can make further transformations of the effective
action 
%if one expands 
by expanding around small frequencies $\omega$.
Our convention for the Fourier transform from time to
frequency domain is
\begin{align}
\tilde{F}(\omega) &= \mathcal{F}(F)
        = \int d \tau \, F(\tau) e^{-i \omega \tau} , \\
F(\tau) &= \mathcal{F}^{-1}(\tilde{F})
        = \frac{1}{2\pi} \int d \omega \, \tilde{F}(\omega)  e^{i \omega \tau} ,
% \delta (\tau) &= \frac{1}{2\pi} \int d\omega e^{\pm i \omega \tau} ,
\end{align}
where $\mathcal{F}$ denotes the Fourier transformation operator and
$\mathcal{F}^{-1}$ its inverse.
Note that we use a tilde to denote Fourier transformed quantities.
% Before going further, we sketch the formal construction of the effective
% action \cite{Goldberger:Rothstein:2006:2}. Prior knowledge of the
% microscopic picture is in principle not needed, however we will see
% that construction will somewhat illuminate what follows here. Let us
% assume that
% the quadrupole dynamics is described by the action
% \be
% S_Q =-\frac{1}{2} \int d\tau Q_{ab} \mathcal O^{ab}{}_{cd}Q{}^{cd} ,
% \label{action_q}
% \ee
% completing the effective action \eqref{eq:effective_action}.
% Variation with respect to $Q$ leads to \eqref{eq:formal_eq_quad},
% solved by 
In the frequency domain, the formal solution for the quadrupole (\ref{Qformal}) reads
\be
\tilde Q{}^{ab} = -\frac{1}{2}\tilde F(\omega) \tilde E^{ab}.
\label{eq:quad_sol}
\ee

\subsection{Relativistic Love numbers}
%We are now going to obtain certain Love numbers from the small frequency
%expansion of the response. 
In this section, we make contact between certain Love numbers and the small
frequency regime of the response function.
In fact, we will eliminate
(integrate out) the quadrupole degrees of freedom from the action.
Inserting the formal solution \eqref{eq:quad_sol} into Eqs.\ \eqref{eq:effective_action}
and \eqref{action_q}, we find that the action turns into
\be
S_{\text{eff}} = \int d\tau \left[ -m +\frac{1}{8} \mathcal F^{-1}(\tilde F
\tilde E_{ab})E^{ab} + \dots \right] .
\label{eq:action_dynamic_q}
\ee
The connection to the relativistic Love numbers and other constants defined in
previous literature  \cite{Hinderer:2007, Damour:Nagar:2009:4, Binnington:Poisson:2009,
Bini:Damour:Faye:2012} becomes apparent if we Taylor-expand the response
\begin{equation}
\tilde{F}(\omega) =  2\mu_2 + i \lambda \omega +  2\mu'_2 \omega^2 +
\Order(\omega^3) ,
\end{equation}
and explicitly perform the inverse Fourier transform in
Eq.\ (\ref{eq:action_dynamic_q}).
The $\lambda$-term is related to absorption \cite{Goldberger:Rothstein:2006:2}.
In this case, the action \eqref{eq:action_dynamic_q} reduces to
\be
S_{\text{eff}} = \int d\tau \left[ -m +\frac{\mu_2}{4} E_{ab} E^{ab}
% + \frac{\lambda}{8}  E_{\mu\nu} \dot E^{\mu\nu}
+ \frac{\mu_2'}{4}  \dot E_{ab} \dot E^{ab} + \dots \right] .
\label{love_action}
\ee
The contribution from $\lambda$ formally turns into a total derivative
and drops out. Indeed, insertions of solutions of equations of motion
into action principles project onto the time-symmetric part of the
dynamics \cite{Galley:2012}, and therefore one must handle
dissipative effects more carefully. (In Ref.\ \cite{Galley:2012}, a classical
variant of the closed time path formalism of quantum field theory was developed
for this purpose.)
Here, $\mu_2$ is the relativistic tidal Love number
\cite{Hinderer:2007, Damour:Nagar:2009:4, Binnington:Poisson:2009}, and $\mu'_2$ represents
the tidal response of the neutron star beyond the
adiabatic approximation \cite{Bini:Damour:Faye:2012}.
It should be noted that $\mu_2$ and $\mu'_2$ are in fact defined by an effective
action of the form (\ref{love_action}) and should be determined numerically through
a matching procedure, just like $\tilde{F}$ (this has not yet been undertaken for $\mu'_2$).

From a historical perspective, the Love numbers of the first and second kind
were introduced in Ref.\ \cite{Love:1911} in the context of tides on Earth.
These are dimensionless numbers describing the response of the
tidal deformation (first Love number or shape Love number) and tidal
potential (second Love number) of an elastic body. In general relativity, there exist
electric-type ($k_l$), magnetic-type ($j_l$), and shape ($h_l$) Love numbers.
The shape and electric-type Love numbers are directly analogous to the
first and second Newtonian Love numbers.
The relation to $\mu_2$ is given by $k_2=3\mu_2 G/2R^5$, where $R$ is the
radius of the star. (The subscript $2$ in $\mu_2$ denotes $l=2$ multipolar contribution and
not the second Love number.)
% Traditionally $k_2$ is known as the ``second Love number'',
% although the subscript $2$ in $k_2$ denotes the quadrupolar contribution
% $l=2$, and it is related to $\mu_2$ by $k_2=3\mu_2 G/2R^5$, where $R$ is the
% radius of the neutron star.
The magnetic-type Love numbers in general relativity were first
introduced in Ref.\ \cite{Damour:Soffel:Xu:1992}. The relativistic
shape Love numbers were considered in Ref.\ \cite{Damour:Nagar:2009:4} for neutron stars
and in Refs.\ \cite{Damour:Lecian:2009, Vega:Poisson:Massey:2011} for black holes.

% To next-to-leading post-Newtonian order it
% is sufficient to consider a coupling of the form
% \begin{equation}\label{SSQaction}
% \frac{1}{\sqrt{- u_{\sigma} u^{\sigma}}} \left(
% 	- \frac{1}{2 m_0} R_{\mu\nu\alpha\beta} S^{\rho\mu} S^{\alpha\beta} u^{\nu} u_{\rho}
% 	+ \frac{C_{ES^2}}{2 m_0} E_{\mu\nu} {S^{\mu}}_{\rho} S^{\rho\nu} \right) \,,
% \end{equation}
% within the action, see \cite{Porto:Rothstein:2008:2} and also \cite{Steinhoff:2011}. Here
% $R_{\mu\nu\alpha\beta}$ is the Riemann tensor, $S^{\alpha\beta}$ is the
% antisymmetric spin tensor, $m_0$ is a constant mass-like parameter, and
% $C_{ES^2}$ is a constant describing the spin-induced quadrupole deformation.
% Higher multipoles can be treated by multiple symmetrized covariant derivatives
% of the curvature tensor, cf.\ equation (19) in \cite{Bailey:Israel:1975}. Notice
% that equation (19) in \cite{Bailey:Israel:1975} gives the connection to Dixon's
% multipole moments \cite{Dixon:1979}. Further notice that though the additions to the action are only accurate
% up to a specific order in some approximation scheme (post-Newtonian approximation,
% multipole approximation, etc.), they are written in a manifestly covariant way.

\subsection{Newtonian limit}
% In purely Newtonian gravity, the tidal Love number of a compact
% object is simply a proportionality constant between the tidal field and the
% resulting multipole moment due to the deformation of the object. In fact in
% Newtonian limit
% \begin{equation}
%  Q_{ab}=\int d^3x \rho(x)\left(x_a x_b-\frac{1}{3}r^2 \delta_{ab}\right),
% \end{equation}
% where $\rho(x)$ is the mass density of the neutron star. 
Finally, for our purpose, we give here the nonrelativistic limit
of the effective action \eqref{eq:effective_action}. First, we note that
\begin{equation}\label{Enonrel}
E_{ab} %\approx R_{0a0b}
\approx \partial_a \partial_b \Phi ,
\end{equation} 
in the Newtonian limit, where $\Phi$ is the Newtonian gravitational potential.
This is obtained from evaluating the electric part of the Weyl tensor to
linear order in the potential with the metric
\bea
ds^2 &=& - \left( 1 + \frac{2 \Phi}{c^2} \right) c^2 dt^2
+ \left(1 + \frac{2\Phi}{c^2} \right)^{-1} dr^2 \nonumber \\
&&+ r^2 d\Omega^2 + \mathcal O\left(\frac{1}{c}\right)^3 ,
\eea
where $x^a = (r, \theta, \varphi)$ and $\Phi$ is the Newtonian potential.
The speed of light $c$ is introduced as a bookkeeping parameter here
and can be removed at the end of the calculation.
The Newtonian limit of $E_{ab}$ is then obtained
straightforwardly by expanding around large $c$.

Next, the proper time is related to the Newtonian absolute time by
\begin{equation}
d\tau = dt \sqrt{- g_{\mu\nu} u^{\mu} u^{\nu}}
        \approx dt \sqrt{1 + 2 \Phi - \dot{\vct{z}}^2},
\end{equation}
where $t$ is the absolute Newtonian time and $\vct{z}$ is the location of the
star.
Finally, the Newtonian limit of Eq.\ \eqref{eq:effective_action} leads to the
effective Newtonian action
\begin{equation}\label{EFTactionNewton}
S_{\text{PP}} \approx \int d t \, \bigg[ - m + \frac{1}{2} m \dot{\vct{z}}^2 -
m \Phi - \frac{1}{2} Q^{ab} \partial_a \partial_b \Phi + \dots \bigg].
\end{equation}
% Remember that (\ref{Enonrel}) holds in the Newtonian limit.

\subsection{Potentials}
We are going to illustrate how the effective action can be applied to
the binary problem and explain the advantages. Effects due to the
internal structure just enter via the quadrupole here. For simplicity, we
assume that the particle described by Eq.\ \ref{EFTactionNewton}) moves
in a fixed gravitational field of a (heavy) point mass $m_{\rm{heavy}}$ located
at the coordinate origin,
\begin{equation}
\Phi = - \frac{G m_{\text{heavy}}}{r} ,
\end {equation}
where $r = |\vct{x}|$.
Insertion into the quadrupole interaction in Eq.\ (\ref{EFTactionNewton})
leads to the monopole-quadrupole
interaction potential
\begin{equation}\label{Vquad}
V_{\text{mono-quad}} = - \frac{3 G m_{ \text{heavy}}}{2 |\vct{z}|^3} Q^{ab} \vct{n}^a \vct{n}^b ,
\end{equation}
where $\vct{n}^a$ is the unit vector pointing toward the quadrupole
particle. Notice that the derivation is rather simple. Further,
the gravitational interaction and the internal dynamics are logically
separated. This can be very advantageous for model building, particularly
in complicated situations.

The derivation of Eq.\ (\ref{Vquad}) is formally valid for large $m_{ \text{heavy}}$
only since we omit backreactions.
However, the result is actually correct for generic binaries.
In general, each body is represented by a copy of Eq.\ (\ref{EFTactionNewton}),
\begin{equation}
S_{\text{binary}} = S_{\text{PP1}} + S_{\text{PP2}} + S_{\Phi} ,
\end{equation}
where the gravitational action is given by
\begin{equation}
S_{\Phi} = \frac{1}{8 \pi G} \int d t \, d^3 x \, \Phi \Delta \Phi .
\end{equation}
Insertion of the solution for $\Phi$ in fact leads to the same result
(\ref{Vquad}) for the monopole-quadrupole potential.
This requires us to drop singular self-interactions, as $\Phi$ is
singular at the position of each particle now.
An extension of the potential to the first post-Newtonian approximation
to general relativity can be found in Ref.\ \cite{Vines:Flanagan:2010}.

In effective field theory parlance, we have in fact just ``integrated out''
the gravitational field $\Phi$, and we will occasionally make use of this
phrase. Indeed, on a classical level, integrating out a field translates
to obtaining its field equations from the action, solving them, and inserting the
solution back into the action (also into the pure-field part $S_{\Phi}$).
The combinatorial aspect suggests to elegantly organize the computation
in a diagrammatic manner {\`a} la Feynman, in particular in a perturbative context.

\section{Action for stellar perturbations\label{stellar_perturbation}}
In this section, we briefly review the variational treatment of linear perturbations
in Newtonian gravity and cast it into the form needed for our investigation.
This will be the starting point for
% the analytic matching to the effective field theory.
the derivation of the effective action later on.
Variational principles for nonradial stellar oscillations
in the Newtonian context were discussed in previous literature
\cite{Chandrasekhar:1964, LyndenBell:Ostriker:1967} and subsequently extended
to tidal excitations in binary systems \cite{Rathore:Broderick:Blandford:2002}.
% A brief outline of the variational principle for perturbations in binary system
% along the lines of 
% \cite{Rathore:Broderick:Blandford:2002} can be found in Appendix \ref{appA}.

In the following, we refer to linear stellar perturbations as the ``full theory.''
This is inaccurate in at least two aspects. First, linear perturbations
are just an approximation to generic perturbations, which can usually only
be tackled by 3-dimensional numerical simulations. Second, on a more fundamental
level, a fluid description for stars is just an approximation, too. Fluids
can be thought of as an effective theory for atomic or subatomic particles,
which themselves may be described by an effective theory of an even more
fundamental theory, and so forth and so on.
This is the effective field theory viewpoint of physical model building, namely
through a tower of effective theories.

\subsection{Variational principle for perturbations\label{var_perturbation}}
% We consider a gravitating ideal fluid configuration described by
% the velocity field of the fluid elements $\vct{u}$,
% usual thermodynamic variables for these fluid elements,
% like mass density $\rho$, pressure $P$,
% and equations of state relating them.
We consider a gravitating ideal fluid configuration described by the following quantities: the velocity field $\mathbf{u}$ of the fluid elements, usual thermodynamic variables, like
mass density $\rho$, pressure $P$, and equations of state relating them.
But we disregard temperature here, which is a good approximation
for neutron stars.
We analyze the system in a perturbative setting, starting from a
background configuration
denoted by an index 0, a first perturbation denoted by an index 1,
and so forth and so on.

For the background, we take a static nonrotating
(spherically symmetric) star in equilibrium, and we restrict to first order
perturbations here.
Then, the unperturbed fluid velocity vanishes, $\vct{u}_0 = 0$.
The other background variables are functions of the radial
coordinate $r$ only and determined by
\begin{gather}
P_0' = - \rho_0 \Phi_0', \label{hydroeq} \\
\Delta \Phi_0 = \frac{1}{r^2} \frac{d ( r^2 \Phi_0' )}{d r} = 4\pi G \rho_0 , \label{backphi}
\end{gather}
where $' = d / d r$.
Notice that these are just the equation of hydrostatic equilibrium and
the Newtonian field equation for spherically symmetric configurations.
Together with a barotropic equation of state $P=P(\rho)$, the system of equations
is closed. In the following, the background variables are not varied in
the action principle but are considered as functions of $r$ given by the
solution to this system of equations. Except in simple cases, the solution
is obtained from numeric integration.

The fundamental variable of stellar perturbations is the
displacement vector field $\vctsym{\xi}$ of the fluid elements.
If $\vct{x'}$ denotes the position of a fluid element in the perturbed star and $\vct{x}$ denotes
the same in the unperturbed background, then
% the Lagrangian displacement $\vctsym{\xi}$ is related to the physical displacement as $\vctsym{\xi} = \vct{x'}-\vct{x}=\delta \vct{x}$.
the physical displacement is $\vctsym{\xi} = \vct{x'}-\vct{x}$.
The perturbation variables can be expressed in terms of the displacement as
\begin{align}
\vct{u}_1 &= \dot{\vctsym{\xi}} , \\
\rho_1 &= - \nabla \cdot (\rho_0 \vctsym{\xi}) , \label{rho1_xi} \\
P_1 &= c_s^2 \rho_1 , \label{P1}
\end{align}
where $c_s$ is the (unperturbed) adiabatic speed of sound given by
$c_s^2=d P_0 / d \rho_0$.
Notice that Eq.\ (\ref{rho1_xi}) follows from an integration of the continuity equation
$\dot{\rho} + \nabla \cdot ( \rho \vct{u} )=0$ (conservation of mass)
together with $\dot{\rho}_0 = 0 = \vct{u}_0$.
We further assume that the perturbation
is caused by some external mass density $\rho^{ext} = \rho^{ext}_1$, which
can represent another close-by star.
% We split the gravitational field perturbation as
% \begin{equation}
% \Phi_1 = \Phi_1^{\text{self}} + \Phi_1^{\text{ext}} ,
% \end{equation}
% into a self-gravitational field $\Phi_1^{\text{self}}$ sourced by $\rho_1$
% (with usual boundary conditions at infinity) and an external field
% $\Phi_1^{\text{ext}}$ with a source outside the body or at infinite distance.
% It is assumed that no external source contributes to $\Phi_0$.
% This is possible thanks to the linearity of the Newtonian gravitational
% field equations.

The Lagrangian of the ``full theory'' (stellar perturbations) can now be
written as a sum of
gravitational, kinetic, star's interior, and coupling parts,
\begin{equation}
L_{\text{full}} = L_{\Phi_1} + L_{\text{kin}} + L_{\text{star}} + L_{\text{coupl}} ,
\end{equation}
where
\begin{align}
L_{\Phi_1} &= \frac{1}{8\pi G} \int d^3 x \, \Phi_1 \Delta \Phi_1 , \quad
L_{\text{kin}} = \frac{1}{2} m \dot{\vct{z}}^2 , \label{LPM} \\
\begin{split}
L_{\text{star}} &= \int d^3 x \bigg[ \frac{1}{2} \rho_0 \dot{\vctsym{\xi}}^2_{\text{COM}}
	- (\rho E)_2
	- \rho_0 \vctsym{\xi} \cdot ( \nabla \Phi_1 + \ddot{\vct{z}} ) \bigg] ,
\end{split}\\
L_{\text{coupl}} &= - \int d^3 x \, \rho_{\text{ext}} ( \Phi_0 + \Phi_1 ) ,
\end{align}
and $E$ is the specific internal energy.
The second-order perturbation of the internal energy
$\rho E(\rho)$ is given by
\begin{equation}
(\rho E)_2 \approx \rho_1 \frac{d E_0}{d \rho_0} \rho_1
        + \frac{1}{2} \rho_0 \frac{d^2 E_0}{d \rho_0^2} \rho_1^2
        = \frac{c_s^2}{2 \rho_0} \rho_1^2 .
\end{equation}
Remember that the first law of thermodynamics reads $d E = - P d \rho^{-1}$
in our case.
Also notice that in order to obtain the first-order perturbation equations,
some second-order contributions must be included in the Lagrangian
(but $\rho_2$ may be omitted).
At this point, the dynamical variables are $\vct{z}$, $\vctsym{\xi}$, and $\Phi_1$.
Here, $\vct{z}$ is the center of mass (COM) of the background solution.
The subscript on $\dot{\vctsym{\xi}}_{\text{COM}}$ indicates that the time
derivative is taken in center-of-mass frame. As no further time derivatives
of fields are present, we assume that all volume integrals are performed
in this frame from now on. Then, the background fields are spherically symmetric
around the origin of integration $\vct{x}=0$.

A derivation of $L_{\text{star}}$ starting from variational fluid dynamics
can be obtained by essentially following Ref.\ \cite{Rathore:Broderick:Blandford:2002}
and is therefore not repeated here.
This derivation assumes a potential for the fluid velocity and a homentropic flow.
One can translate Ref.\ \cite[Eqs.\ (17), (18), and (19)]{Rathore:Broderick:Blandford:2002}
to our formulation by making use of the background solutions and expressing the
perturbation variables in terms of $\vctsym{\xi}$.
However, the contributions to $L_{\text{star}}$ have simple interpretations.
The first term is just
the kinetic energy of the perturbation. The second term comes from the change
of the internal energy [$(\rho E)_1$ is canceled by insertion of the
background solutions].
Finally, the last term is the coupling of the perturbation
to gravitation and to the fictitious force due to the overall acceleration
$\ddot{\vct{z}}$ of the star.

% Using Eq (\ref{rho1_xi}), we note that
% \begin{equation}
% - c_s^2 \nabla \cdot (\rho_0 \vctsym{\xi}) = \frac{d P_0}{d \rho_0} \rho_1 = P_1 ,
% \end{equation}
% represents forces due to pressure perturbations. Finally, the variation of the above action leads
% to two perturbation equations,
% \begin{align}
% \begin{split}
% \rho_0 \ddot{\vctsym{\xi}} &= - \rho_0 \nabla \left( \frac{P_1}{\rho_0} + \Psi_1 \right) + \vct{f}_{\text{tot}} , \\
% &= - \rho_0 \nabla \left( - \frac{c_s^2}{\rho_0} \nabla \cdot (\rho_0 \vctsym{\xi})
% 	+ \Psi_1 + \Phi + \vct{x} \cdot \ddot{\vct{R}}_{*} \right) ,
% \end{split}\\
% \Delta \Psi_1 &= 4\pi G \rho_1 = - 4\pi G \nabla \cdot (\rho_0 \vctsym{\xi}) ,
% \end{align}
% where $P_1$ is understood as an abbreviation for (\ref{P1}). We will compare
% these equations to their relativistic counterparts in
% \cite{Lindblom:Mendell:Ipser:1997} later on.

In the following, we are actually not going to substitute
$\rho_{\text{ext}}$ by another compact object. Instead, we view it
as an external ``current,'' similar as for generating functionals
in quantum field theory. The important observation is that
\begin{equation}\label{PhiLegendre}
\Phi = \Phi_0 + \Phi_1 = - \frac{ \delta L_{\text{full}} }{ \delta \rho_{\text{ext}} } .
\end{equation}
This enables us to recover the gravitational field from the Lagrangian
even after it was integrated out.

\subsection{Normal modes}
To define the normal oscillation modes of the star, we need to
integrate out the gravitational field perturbation.
The formal solution to its field equation reads
\begin{equation}
\Phi_1 = 4\pi G \Delta^{-1} \left[ \rho_1 + \rho_{\text{ext}} \right] ,
\end{equation}
where the inverse Laplacian $\Delta^{-1}$ is defined
% as usual by an integral involving the gravitational Green function
for usual boundary conditions at spatial infinity.
After insertion into the Lagrangian, we reorder the Lagrangian into
normal mode (NM), interaction, and pure external parts (keeping only the kinetic part untouched),
\begin{equation}\label{Lfull}
L_{\text{full}} = L_{\text{kin}} + L_{\text{NM}} + L_{\text{int}} + L_{\text{ext}} ,
\end{equation}
where
\begin{align}
L_{\text{NM}} &= \int d^3 x \, \frac{\rho_0}{2} \bigg[ \dot{\vctsym{\xi}}^2_{\text{COM}}
	- \vctsym{\xi} \cdot \mathcal{D} \vctsym{\xi} \bigg] , \\
L_{\text{int}} &= - \int d^3 x \, [ \rho_0 \Phi_{\text{ext}}
        + \rho_1 ( \Phi_{\text{ext}} + \vct{x} \cdot \ddot{\vct{z}} ) ] , \label{Lint} \\
L_{\text{ext}} &= - \frac{1}{2} \int d^3 x \, \rho_{\text{ext}} \Phi_{\text{ext}} , \label{Lext}
\end{align}
with the abbreviation $\Phi_{\text{ext}} = 4\pi G \Delta^{-1} \rho_{\text{ext}}$
and the linear (nonlocal, integro-differential) operator $\mathcal{D}$ defined by
\begin{equation}
\mathcal{D} \vctsym{\xi} = - \nabla \left\{ \left[ \frac{c_s^2}{\rho_0} + 4\pi G \Delta^{-1} \right] \nabla \cdot (\rho_0 \vctsym{\xi}) \right\} .
\end{equation}

It is easy to see that $\mathcal{D}$ is Hermitian with respect to the
compact integration measure $d m_0 = \rho_0 d^3 x$, which was first recognized by
Chandrasekhar \cite{Chandrasekhar:1964}.
% Variation of the action now leads to
% \begin{equation}
% \ddot{\vctsym{\xi}} = \mathcal{D} \vctsym{\xi} - \nabla \left( \Phi + \vct{x} \cdot \ddot{\vct{R}}_{*} \right) .
% \end{equation}
This fact guarantees that it possesses a complete set of
eigenfunctions $\vctsym{\xi}^{\text{NM}}_{nlm}$ with real eigenvalues $\omega_{nl}^2$,
\begin{equation}
\mathcal{D} \vctsym{\xi}^{\text{NM}}_{nlm} = \omega_{nl}^2 \vctsym{\xi}^{\text{NM}}_{nlm} ,
\end{equation}
which are orthogonal and can be normalized,
\begin{equation}
\int d^3 x \, \rho_0 \vctsym{\xi}^{\text{NM} \, \dagger}_{n'l'm} \vctsym{\xi}^{\text{NM}}_{nlm}
	= \delta_{n'n} \delta_{l'l} \delta_{m'm} . \label{xinorm}
\end{equation}
% where the prefactor $m$ is the mass of the star and $R$ its radius.
This normalization is convenient for calculations but implies unusual units
for $\vctsym{\xi}^{\text{NM}}_{nlm}$.

This defines the normal modes of the star.
The modes are discrete due to the compact integration measure.
As $\mathcal{D}$ respects rotational symmetry (in the center-of-mass system),
one can label the normal modes
by the usual angular momentum ``quantum'' numbers $l$ and $m$.
(When $m$ is not used as an index, it always refers to the mass.) In
particular, the perturbation quantities can be expanded over the relevant
(scalar, vector,...) spherical harmonics and inherit the associated angular
symmetries.
Notice that $\omega_{nl}$ is independent of $m$.
% that is \cite{Regge:Wheeler:1957, Thorne:1980}
% \begin{align}
% \vctsym{\xi}^{\text{NM}}_{nlm} &= \xi^{\text{RNM}}_{nl}(r) \vct{Y}^{R,lm}(\theta, \phi)
% 	+ \xi^{\text{ENM}}_{nl}(r) \vct{Y}^{E,lm}(\theta, \phi) \nonumber \nl
% 	+ \xi^{\text{BNM}}_{nl}(r) \vct{Y}^{B,lm}(\theta, \phi) .
% \end{align}
The label $n$ denotes the radial structure of the modes.
Usually these are given as a letter ($f$, $p$, $g$, etc), labeling a general class to
which the modes belong by considering the primary restoring mechanism (corresponding
to fundamental, pressure, gravity modes), together with a number counting
its radial nodes (analogous to the overtone number for a vibrating string).
% The three different parts in the above set of equations correspond to radial (R), electric-type (E), and magnetic type
% (B) modes with corresponding orthonormal vector spherical harmonics \cite[Eq 2.15]{Thorne:1980}
% \begin{align}
% \vct{Y}^{E,lm}(\theta, \phi) &= \frac{1}{\sqrt{l (l+1)}} r \nabla Y^{lm}(\theta, \phi) , \\
% \vct{Y}^{B,lm}(\theta, \phi) &= \vct{n} \times \vct{Y}^{E,lm}(\theta, \phi) , \\
% \vct{Y}^{R,lm}(\theta, \phi) &= \vct{n} Y^{lm}(\theta, \phi) ,
% \end{align}
% Here $Y^{lm}$ are the usual scalar spherical
% harmonics and $\vct{n} = \vct{x} / r$.

\subsection{Amplitude formulation}
A general physical displacement $\vctsym{\xi}$ can now be decomposed into a sum
of discrete normal modes $\vctsym{\xi}^{\text{NM}}_{nlm}$ with corresponding
time-dependent amplitudes $A_{nlm}$,
\begin{equation}
\vctsym{\xi} = \sum_{nlm} A_{nlm}(t) \vctsym{\xi}^{\text{NM}}_{nlm}(\vct{x}) ,
\end{equation}
thanks to the completeness of the $\vctsym{\xi}^{\text{NM}}_{nlm}$.
The reality condition $\vctsym{\xi}^{*} = \vctsym{\xi}$ leads to
\begin{equation}
A_{nlm}^{*} = (-1)^{m} A_{nl \, -m} ,
\end{equation}
which follows from the property $Y^{lm *} = (-1)^{m} Y^{l \, -m}$
of the spherical harmonics.
% under the assumption that the $r$-dependent functions are real.
Additionally, according to Eq.\ (\ref{rho1_xi}), it holds
\begin{equation}
\rho_1 = \sum_{nlm} A_{nlm} \rho^{\text{NM}}_{nlm} , \quad
\rho^{\text{NM}}_{nlm} := - \nabla \cdot (\rho_0 \vctsym{\xi}^{\text{NM}}_{nlm} ) .
\label{rhodecomp}
\end{equation}
Notice that we can separate the angular dependence as
\begin{equation}
\rho^{\text{NM}}_{nlm} = \rho^{\text{NM}}_{nl}(r) Y^{l m}(\Omega) , \label{rhoseparate}
\end{equation}
and that the radial part $\rho^{\text{NM}}_{nl}$ must be a real function.
This property allows the use of the orthogonality properties of spherical harmonics
under angular integration $d \Omega$, which will become important below.

The normal mode and interaction Lagrangians turn into
\begin{align}
L_{\text{NM}} &= \sum_{nlm} \frac{1}{2} \left[ | \dot{A}_{nlm} |^2
	- \omega_{nl}^2 | A_{nlm} |^2 \right] , \label{LNM} \\
L_{\text{int}} &= \sum_{nlm} A_{nlm} f_{nlm}^{*}
        - \int d^3 x \, \rho_0 \Phi_{\text{ext}} ,
% &= \sum_{nlm} \frac{(-1)^{m}}{2} \bigg[ \dot{A}_{nlm} \dot{A}_{nl \, -m}
% 	- \omega_{nl}^2 A_{nlm} A_{nl \, -m} \bigg] ,
\end{align}
where the force term is given by
\begin{equation}
f_{nlm} = - \int d^3x \, \rho^{\text{NM} \, *}_{nlm} ( \Phi_{\text{ext}} + \vct{x} \cdot \ddot{\vct{z}} ) .
\label{overlap}
\end{equation}
The discrete amplitudes $A_{nlm}$ now take on the role of dynamic variables
in place of $\vctsym{\xi}$.
Their equations of motion have the form of a forced harmonic oscillator,
\begin{equation}
\ddot{A}_{nlm} + \omega_{nl}^2 A_{nlm} = f_{nlm} , \label{eq:amplitude_evolution}
\end{equation}
where we used $f_{nlm}^{*} = (-1)^{m} f_{nl \, -m}$ after variation of
the action.
This amplitude formulation is the starting point for investigations
in Refs.\ \cite{Gingold:Monaghan:1980, Alexander:1987,
Rathore:Blandford:Broderick:2004, Flanagan:Racine:2007}; see also
Refs.\ \cite{Chandrasekhar:1964, LyndenBell:Ostriker:1967, Press:Teukolsky:1977,
Rathore:2005}.
% One can phenomenologically add dissipation by using damped oscillators.

\section{Further developments}\label{devlpmnt}
So far we have used the angular momentum numbers $l$ and $m$ to characterize the perturbation.
Alternatively, one can transform the spherical harmonics to Cartesian
tensors, specifically STF tensors.
This is useful because the multipoles appearing in the effective action
are STF tensors. We are also going to relate the multipoles and the
amplitudes in this section.
A brief account on the STF formalism is given in Appendix \ref{STF}.

\subsection{STF basis}
When written in terms of STF tensors, the expressions above are essentially the same, but the
index $m$ is replaced by STF tensor indices of rank $l$.
This is essentially just a basis change of a vector space: The index $m$ can take on
values between $-l$ and $l$ and thus labels components of a $2l+1$-dimensional
vector, while the independent components of a STF rank-$l$
tensor exactly matches $2l+1$. This is not a coincidence but deeply rooted in properties of
rotation group representations: STF tensors transform irreducibly under
rotations, just like spherical harmonics.

We denote the basis transformation matrix by $\mathcal{Y}$, such that it holds
\begin{align}
Y^{00} &= \mathcal{Y}^{00} , \\
Y^{1m} &= \mathcal{Y}^{1m}_{k_1} n^{k_1} , \\
Y^{2m} &= \mathcal{Y}^{1m}_{k_1 k_2} n^{k_1} n^{k_2} , \\
&\vdots \nonumber \\
Y^{lm} &= \mathcal{Y}^{lm}_{K_l} \hat{n}^{K_l} ,
\end{align}
where $n^k = n^k(\Omega)$ is a unit vector. In the last line, we
adopted a multi-index $K_l = \{ k_1, k_2, \dots, k_l \}$ and the abbreviation
\begin{equation}
\hat{n}^{K_l} = [ n^{k_1} n^{k_2} \dots n^{k_l} ]^{\text{STF}} .
\end{equation}
For instance, $\hat{n}^{ij} = n^{i} n^{j} - \delta^{ij}/3$.
Because of the use of Cartesian multipoles in the effective action,
it is best to transform the oscillator amplitudes and force
integrals to the STF basis from now on,
which is denoted by a hat,
\begin{align}
\hat{A}_{n K_l} &= \sum_m A_{nlm} N_l \mathcal{Y}^{lm}_{K_l} , \\
\hat{f}_{n K_l} &= \sum_m f_{nlm} N_l \mathcal{Y}^{lm}_{K_l} , \label{fSTF}
\end{align}
where the normalization factor $N_l$ is defined and explained in
Appendix \ref{STF}.
% The relevant Lagrangians translate into
% \begin{align}
% L_{\text{NM}} &= \sum_{nl} \frac{1}{2} \left[ ( \hat{A}_{n K_l} )^2
% 	- \omega_{nl}^2 ( \hat{A}_{n K_l} )^2 \right] , \\
% L_{\text{int}} &= \sum_{nl} \hat{A}_{n K_l} \hat{f}_{n K_l}
%         - \int d^3 x \, \rho_0 \Phi_{\text{ext}} , \label{Lintamp}
% \end{align}
% which is expected from a simple change of basis.

An immediate consequence is that the acceleration term in
Eq.\ \ref{overlap}) contributes only for $l=1$, which is due to
$\vct{x} = r \vct{n} \sim Y^{1m}$ and the orthogonality of the
spherical harmonics. We are going to analyze the contributions of
$\Phi_{\text{ext}}$ in a similar manner now.

\subsection{Multipoles and overlap integrals}
The Cartesian mass multipole moments of the perturbation are defined by
\begin{equation}
Q^{K_l} = \int d^3 x \, \rho_1 r^l \hat{n}^{K_l} . \label{Qdef}
\end{equation}
It should be emphasized that the multipoles $Q^{K_l}$ are not the fundamental dynamical
variables of the theory, but are composed of the mode amplitudes.
This becomes explicit by making use of Eqs.\ (\ref{rhodecomp}) and
(\ref{rhoseparate}) in Eq.\ (\ref{Qdef}),
\begin{align}
Q^{K_l} &= \sum_{nl'm} A_{nl'm} \int d r \, r^{l+2} \rho^{\text{NM}}_{nl'}(r)
        \int d \Omega \, Y_{l'm} \hat{n}^{K_l} , \nonumber \\
&= \sum_n I_{nl} \hat{A}_{n K_l} , \label{composed}
\end{align}
where
\begin{equation}
I_{nl} = N_l \int d r \, r^{l+2} \rho^{\text{NM}}_{nl}(r) . \label{Ioverlap}
\end{equation}
The angular integration is given by Eq.\ (\ref{Iangular}).
However, the result is very plausible. Only modes with angular momentum
number $l$ can contribute to the $2l$-pole $Q^{K_l}$.

The quantity $I_{nl}$ is better known as the overlap integral.
(We included the factor $N_l$ in contrast to other publications, see
Appendix \ref{overlapconvention}.)
It describes to which extend an external field excites the mode.
% This becomes clear by analyzing the interaction Lagrangian in STF basis,
% \begin{equation}
% L_{\text{int}} = \sum_{nl} \hat{A}_{nl K_l} \hat{f}_{nl K_l} , \label{Lintamp} .
% \end{equation}
% Comparing (\ref{Lintamp}) to (\ref{LintEQ}) and (\ref{composed}) we see that
To understand this, we need to analyze the force terms $\hat{f}_{nl K_l}$
defined by Eq.\ (\ref{fSTF}). Here, the external field enters through Eq.\ (\ref{overlap}).
As the source of $\Phi_{\text{ext}}$ is located outside of the star,
we can expand it by a Taylor series around the center $\vct{z}$,
\begin{equation}
\begin{split}
\Phi_{\text{ext}} &= \Phi_{\text{ext}}(\vct{z}) + x^i (\partial_i \Phi_{\text{ext}})(\vct{z}) \nl
	+ \frac{1}{2} x^i x^j (\partial_i \partial_j \Phi_{\text{ext}})(\vct{z}) + \dots .
\end{split}
\end{equation}
Because of the Laplace equation $\Delta\Phi_{\text{ext}} = 0$ valid inside the star,
the traces of the multiple partial derivatives can be removed. This results in
\begin{equation}\label{TaylorPhi}
\Phi_{\text{ext}} = \sum_l \frac{1}{l!} r^l \hat{n}^{K_l} (\partial_{K_l} \Phi_{\text{ext}})(\vct{z}) .
\end{equation}
It is important that $\hat{n}^{K_l}$ is orthogonal to all spherical
harmonics with angular momentum number different from $l$.

We insert Eq.\ (\ref{TaylorPhi}) into the integral in Eq.\ (\ref{overlap}) to arrive at
\begin{align}
\hat{f}_{nl K_l} = - \frac{I_{n l}}{l!} (\hat{\partial}_{K_l} \Phi_{\text{ext}})(\vct{z}) ,
\label{eq:def_f}
\end{align}
for $l \neq 1$, where we made heavy use of the formulas in Appendix \ref{STF}.
The hat over $\partial_{K_l}$ again denotes STF projection. For $l=1$ it holds
\begin{equation}
\hat{f}_{n 1 k} = - I_{n 1} [ (\partial_k \Phi_{\text{ext}})(\vct{z}) + \ddot{\vct{z}} ]
        \approx 0 , \label{fdipole}
\end{equation}
where we used the leading-order equation of motion for $\vct{z}$. This means
that dipole oscillations are not excited in binary systems (by linear
perturbations).

% We insert (\ref{TaylorPhi}) into the integrals in (\ref{overlap})
% and find for $l \neq 1$
% \begin{align}
% \begin{split}
% \hat{f}_{nl K_l} &= - \sum_{l' m} \frac{N_l}{l'!} \mathcal{Y}^{lm}_{K_l}
%         (\partial_{K_{l'}} \Phi_{\text{ext}})(\vct{z}) \nl
%         \times \int d r \, r^{l'+2} \rho^{\text{NM}}_{nl}(r)
%         \int d \Omega \, Y_{lm}^{*} \hat{n}^{K_{l'}} ,
% \end{split}\\
%         &= - \frac{I_{n l}}{l!} (\hat{\partial}_{K_l} \Phi_{\text{ext}})(\vct{z}) ,
% \label{eq:def_f}
% \end{align}
% where
% \begin{equation}
% I_{nl} = N_l \int d r \, r^{l+2} \rho^{\text{NM}}_{nl}(r) . \label{Ioverlap}
% \end{equation}
% The angular integration is given by (\ref{Iangular}).
% However, the result is very plausible. Only modes with angular momentum
% number $l$ can contribute to force $\hat{f}_{nl K_l}$ on the $l$-amplitudes.
% The hat over $\partial_{K_l}$ again denotes STF projection.

It should be noted that the theory is invariant under $A_{nlm} \rightarrow - A_{nlm}$
and $\vctsym{\xi}^{\text{NM}}_{nlm} \rightarrow - \vctsym{\xi}^{\text{NM}}_{nlm}$.
This also implies that $\rho^{\text{NM}}_{nl} \rightarrow - \rho^{\text{NM}}_{nl}$
and consequently $I_{n l} \rightarrow - I_{n l}$. Therefore, we can always make
$I_{n l}$ positive, $I_{n l} \geq 0$. We assume this to be the case.

\section{Derivation of the effective action\label{micro}}
The standard formalism for tidal interactions was cast into an action
approach in the last section. Based on this result, the effective action
can be derived in an elegant manner.
This action will indeed be of the form (\ref{EFTactionNewton}), which was obtained from an
effective action in general relativity by taking the nonrelativistic limit.

\subsection{Multipole expansion}
We are going to show that the interaction Lagrangian can be written in the
same form as the effective action (\ref{EFTactionNewton}).
% As the source of $\Phi_{\text{ext}}$ is located outside of the star,
% we can expand it in a Taylor series,
% \begin{equation}
% \begin{split}
% \Phi_{\text{ext}} &= \Phi_{\text{ext}}(\vct{z}) + x^i (\partial_i \Phi_{\text{ext}})(\vct{z}) \nl
% 	+ \frac{1}{2} x^i x^j (\partial_i \partial_j \Phi_{\text{ext}})(\vct{z}) + \dots .
% \end{split}
% \end{equation}
% Due to the Laplace equation $\Delta\Phi_{\text{ext}} = 0$ valid inside the star,
% the traces of the multiple partial derivatives can be removed. This results in
% \begin{equation}
% \Phi_{\text{ext}} = \sum_l \frac{1}{l!} r^l \hat{n}^{K_l} (\partial_{K_l} \Phi_{\text{ext}})(\vct{z}) .
% \end{equation}
% It is important that $\hat{n}^{K_l}$ is orthogonal to all spherical
% harmonics with angular momentum number different from $l$.
For this purpose, we insert Eq.\ (\ref{TaylorPhi}) into the interaction Lagrangian (\ref{Lint}).
Because of the spherical symmetry of $\rho_0 \sim Y^{00}$, it is easy to see that
\begin{equation}
- \int d^3 x \, \rho_0 \Phi_{\text{ext}} = - m \Phi_{\text{ext}}(\vct{z}) ,
\end{equation}
where
\begin{equation}
m = \int d^3 x \, \rho_0 = \text{const} .
\end{equation}
Furthermore, it holds
\begin{equation}
% - \int d^3 x \, \rho_1 \vct{x} \cdot \ddot{\vct{z}} = - Q^i \ddot{z}^i, \\
- \int d^3 x \, \rho_1 \Phi_{\text{ext}} = - \sum_l \frac{1}{l!} Q^{K_l} (\partial_{K_l} \Phi_{\text{ext}})(\vct{z}) ,
\end{equation}
where the $Q^{K_l}$ are just the usual mass multipoles of the perturbation
(\ref{Qdef}). We assume that
\begin{gather}
Q = \int d^3x \, \rho_1 = 0 , 
\end{gather}
which implies that the mass of the compact object is not modified by the perturbation.
Further, the terms from Eq.\ (\ref{Lint}) involving the mass dipole $Q^i$ combine to
$- [ (\partial_i \Phi_{\text{ext}})(\vct{z}) + \ddot{z}^i ] Q^i$, which vanishes by virtue
of the leading-order equation of motion for $\vct{z}$ and is of higher order.
That is, this term can be removed at the Lagrangian level by a redefinition
of $\vct{z}$, see Ref.\ \cite{Damour:Schafer:1991},
which can produce further terms only at quadratic perturbation order.
% The dipole term $\vct{D}$ is related to the
% spin component $S^{0i}$, and we will ignore it for now. In fact, it is
% possible to set $\vct{D}=0$ by properly choosing the coordinate system
% (generally not the center of mass frame of the background).
% choice for the reference point $R_{*}$ within the object one can modify
% $\vct{D}$ to virtually any desired value, e.g., $\vct{D} = 0$. However, one
% should keep in mind that $\vct{x} = 0$ is then in general not the center of mass
% of the background solution any more. 
% This has to be taken into account when
% determining the multipole moments or the NM of the star.

The result for the interaction Lagrangian (\ref{Lint}) is
\begin{equation}
L_{\text{int}} = - m \Phi_{\text{ext}}(\vct{z})
- \sum_{l \geq 2} \frac{1}{l!} Q^{K_l} (\partial_{K_l} \Phi_{\text{ext}})(\vct{z}) . \label{LintEQ}
\end{equation}
The applied method of deriving the multipole expansion was suggested and
developed further in Ref.\ \cite{Ross:2012} in the relativistic case.
Notice that we have localized the Lagrangian on the center-of-mass
position $\vct{z}$, which means that we represent the extended
object by a point particle comprising various multipole moments
$Q^{K_l}$. By cutting off the multipole summation at some value
of $l$, we can neglect effects of small-scale structure in a controlled
manner. That is, the summation in Eq.\ (\ref{LintEQ}) is a sum over
\emph{interaction energies} and one only needs to include the terms relevant
for the desired accuracy. Here a clear separation of scales is
crucial.

Remember that the multipoles $Q^{K_l}$ are not the fundamental dynamical
variables of the theory but are composed of the mode amplitudes;
see Eq.\ (\ref{composed}). The most explicit form of the interaction
Lagrangian thus reads
\begin{equation}
L_{\text{int}} = - m \Phi_{\text{ext}}(\vct{z})
- \sum_{n, l \geq 2} \frac{I_{nl}}{l!} (\partial_{K_l} \Phi_{\text{ext}})(\vct{z}) \hat{A}_{n K_l} .
\end{equation}
From this result, we can also easily obtain $\hat{f}_{nl K_l}$ as
the coefficient of $\hat{A}_{n K_l}$, cf.\ Eq.\ (\ref{eq:def_f}).
The kinematic terms for the STF amplitudes $\hat{A}_{n K_l}$ simply read
\begin{align}
L_{\text{NM}} &= \sum_{nl} \frac{1}{2} \left[ ( \dot{\hat{A}}_{n K_l} )^2
	- \omega_{nl}^2 ( \hat{A}_{n K_l} )^2 \right] , \label{LNMSTF}
\end{align}
which is clear from the unitarity of the transformation to
the STF basis.

% As an alternative we perform a multipole decomposition of $\rho_1$ and
% $\vctsym{\xi}$,
% \begin{align}
% \rho_1 &= \left[ - D^i \partial_i + \frac{1}{2} Q^{ij} \partial_i \partial_j - \dots \right] \delta_{*} , \\
% \rho_0 \xi^i &= \left[ D^i - \frac{1}{2} Q^{ij} \partial_j - \dots \right] \delta_{*},
% \end{align}
% where $\delta_{*} = \delta(\vct{x} - \vct{R}_{*})$ and (\ref{rho1_xi}) was used.
% It is then easy to derive (\ref{LintEQ}) from these relations.

\subsection{Effective action\label{EFT}}
The final step is to construct the effective action according to its
definition in quantum field theory. That is, we reintroduce the
gravitational field as a dynamical variable through a Legendre
transformation using Eq.\ (\ref{PhiLegendre}). (This should not be
confused with the Wilsonian definition of the effective action.)

The effective Lagrangian is the Legendre transform
\begin{equation}\label{legendre}
L_{\text{eff}} = \int d^3 x \, \rho_{\text{ext}} \Phi + L_{\text{full}} ,
\end{equation}
where the solution for $\rho_{\text{ext}}$ from Eq.\ (\ref{PhiLegendre}) must be inserted.
[The unusual sign in Eq.\ (\ref{legendre}) is consistent with Eq.\ (\ref{PhiLegendre}).]
Recalling Eqs.\ (\ref{Lfull}), (\ref{Lext}), and (\ref{LintEQ}) and the
abbreviation $\Phi_{\text{ext}} = 4\pi G \Delta^{-1} \rho_{\text{ext}}$,
we can evaluate Eq.\ (\ref{PhiLegendre}) and solve it for $\rho_{\text{ext}}$,
\begin{equation}\label{rhoLengendre}
\rho_{\text{ext}} = \frac{1}{4\pi G} \Delta \Phi - m \delta
        - \sum_{l \geq 2} \frac{(-1)^l}{l!} Q^{K_l} \partial_{K_l} \delta ,
\end{equation}
where $\delta = \delta(\vct{x} - \vct{z})$ and $\vct{x}$ is the
field coordinate in an inertial frame now. Also notice that
the Legendre transformation (\ref{legendre}) involves the full field $\Phi$,
while Eq.\ (\ref{Lext}) comprises the external part only. As an intermediate step,
we notice that
\begin{equation}
L_{\text{eff}} = L_{\text{kin}} + L_{\text{NM}}
        + \frac{1}{2} \int d^3 x \, \rho_{\text{ext}} \, 4\pi G \Delta^{-1} \rho_{\text{ext}} ,
\end{equation}
where $L_{\text{NM}}$ is still given by Eq.\ (\ref{LNM}) or (\ref{LNMSTF}).
Here, we still need to insert Eq.\ (\ref{rhoLengendre}), which produces
singular self-interactions, like $\delta \Delta^{-1} \delta$.
These are simply dropped here.
The physical origin of these singularities is the inability of the
multipole expansion to reproduce the gravitational field inside the
body. The situation is completely analogous to the electrostatic
energy of a charge distribution in the point-charge (monopole) limit
and discussed in many textbooks.
Ignoring these self-interactions, the result of the Legendre transformation
reads
\begin{equation}\label{Left}
L_{\text{eff}} = L_{\Phi} + L_{\text{NM}} + L_{\text{PP}} ,
\end{equation}
where
\begin{align}
L_{\text{PP}} &= \frac{1}{2} m \dot{\vct{z}}^2 - m \Phi
        - \sum_{l \geq 2} \frac{1}{l!} Q^{K_l} \partial_{K_l} \Phi , \label{LPP} \\
L_{\Phi} &= \frac{1}{8\pi G} \int d^3 x \, \Phi \Delta \Phi .
\end{align}
The argument $\vct{z}$ of the fields was omitted in $L_{\text{PP}}$
for simplicity.
The general relativistic (covariant) generalization of $L_{\text{PP}}$ is given by
Ref.\ \cite[Eq.\ (20)]{Goldberger:Rothstein:2006:2} to quadrupole order $l=2$,
see also Sec.\ \ref{effaction} here, and our generic result (\ref{LPP})
can be compared immediately to Ref.\ \cite[Eq.\ (1)]{Goldberger:Ross:2009}.

It is interesting to interpret the derivation given in the present section
in the context of Wilson's effective action. In the standard construction,
the field $\Phi$ is split into short-scale (ultraviolet) and long-scale (infrared)
parts. This is best done in spatial Fourier domain $\vct{k}$, where the operator
$\Delta^{-1}$ is a local one. Then, the procedure is as follows:
\begin{enumerate}
\item The first step is to integrate out
only the ultraviolet part of $\Phi$.
\item Next, $\vct{k}$ is rescaled such that
the now vacant ultraviolet regime is repopulated by the infrared contributions.
At this point, one has basically zoomed out and views the system at a larger
scale.
\item In a final step, the dynamical variables are renormalized in order
to recover the original normalization of the kinematic terms in the action.
\end{enumerate}
Here, the procedure is different but essentially analogous:
\begin{enumerate}
\item The field is first integrated out entirely (not just the
ultraviolet
part). This is beneficial for defining the operator $\mathcal{D}$ and
subsequent definition of the oscillation modes.
\item The subsequent Taylor or multipole
expansion shrinks the object to a point, which projects onto the infrared
scales larger than the size of the object (remember that the multipole approximation breaks
down in the interior). This basically zooms out, but without the need for an
explicit rescaling.
\item Finally, in the course of Legendre transformation to the
effective action, divergent terms were dropped. This corresponds to an
implicit renormalization of the dynamical variables.
\end{enumerate}
The similarities to the standard construction are obvious.

There exists a shortcut to Eq.\ (\ref{LPP}), which is characterized as the
continuum effective field theory in Ref.\ \cite{Georgi:1994}. It is an intuitive assumption
that $L_{\text{PP}}$ should be local. Further, invariance under rotations is even
strictly required. It is also possible to absorb certain terms
by variable redefinitions. Such considerations allow one to restrict the
most generic possible form of $L_{\text{PP}}$ considerably. This is how
the general relativistic version of Eq.\ (\ref{LPP}) was constructed in
Re.\ \cite{Goldberger:Ross:2009}; see also Refs.\ \cite{Goldberger:Rothstein:2006,
Goldberger:Rothstein:2006:2}. The idea is then to fix the remaining arbitrariness
of the Lagrangian (here given by the coefficients $Q^{K_l}$) through a
matching against the full theory. This approach will be followed in the
next section. It is usually simpler than an explicit derivation of the
effective action, especially for nonlinear theories like general relativity.
In the present Newtonian case, the advantage is not incredible but still illustrative.

\section{Analytic matching\label{match}}
In this section, we assume that the generic form of the effective
action was constructed, e.g., from symmetry arguments. In our case,
this leads to Eq.\ (\ref{EFTactionNewton}) at quadrupole order or
Eq.\ (\ref{LPP}) to all multipole orders. The precise arguments are given in
Refs.\ \cite{Goldberger:Rothstein:2006, Goldberger:Rothstein:2006:2, Goldberger:Ross:2009},
even for the relativistic case, and are not repeated here.
However, in this section, we pretend that
we are completely uninformed about the fact that the multipoles
are composed of mode amplitudes (\ref{composed}) and about the
Lagrangian $L_{NM}$ for them. Instead, we establish the connection
between the constructed effective action and the full theory from
Sec.\ \ref{stellar_perturbation} through a matching procedure.
The result will be the response function introduced in
Sec.\ \ref{int_response}.

\subsection{Matching condition}
We are going to fit together the gravitational field $\Phi$ predicted by the
effective theory (\ref{Left}) to the desired one of the full theory
(\ref{Lfull}). Without loss of generality, we assume $\vct{z} = 0$ from
now on. The matching condition can be formulated as
\begin{equation}
\frac{\delta L_{\text{eff}}}{\delta \rho_{\text{ext}}}
        = - \Phi = \frac{\delta L_{\text{full}}}{\delta \rho_{\text{ext}}} ,
\end{equation}
which should hold at large scales, i.e., for $r \gg R$, where $R$
is the radius of the compact object.

We first evaluate the left-hand side,
\begin{equation}
\Phi^{\text{eff}} = 4\pi G \Delta^{-1} \left[ \rho_{\text{ext}} + m \delta
        + \sum_{l \geq 2} \frac{(-1)^l}{l!} Q^{K_l} \partial_{K_l} \delta \right] ,
\end{equation}
which is just the inverse of Eq.\ (\ref{rhoLengendre}).
% We will restrict to the case $l=2$ for clarity from now on.
% The generalization of the results for higher multipoles
% can be found in Appendix \ref{higher}.
% Using $4\pi G \Delta^{-1} \delta = - r^{-1}$ and
% \be
% \partial_i\partial_j \frac{1}{r} = \frac{3 n^i n^j - \delta_{ij}}{r^3} ,
% \ee
% we arrive at to the following potential generated by a monopole $m$
% and a quadrupole $Q^{ij}$
% \begin{equation}
% \Phi = 4\pi G \Delta^{-1} \rho_{\text{ext}} - \frac{G m}{r}
%         - \frac{3 G}{2} Q^{ij} \frac{n^i n^j}{r^3} + \dots .
% \end{equation}
Using $4\pi \Delta^{-1} \delta = - r^{-1}$ and Eq.\ (\ref{multipotential}), we we
arrive at to the potential generated by the multipoles,
\begin{equation}\label{PhiEFT}
\Phi^{\text{eff}} = - \frac{G m}{r}
        - \sum_{l \geq 2} \frac{G (2l-1)!!}{l!} Q^{K_l} \frac{\hat{n}_{K_l}}{r^{l+1}}
        + \Phi_{\text{ext}} .
\end{equation}
Remember that we act as if the composition of the $Q^{K_l}$,
Eq.\ (\ref{composed}), and the form of $L_{\text{NM}}$ in the effective
theory are unknown to us.

On the other hand, in the full theory, it holds
\begin{equation}
\Phi^{\text{full}} = - \frac{\delta L_{\text{full}}}{\delta \rho_{\text{ext}}}
= 4\pi G \Delta^{-1} \left[ \rho_0 + \rho_1 \right]
        + \Phi_{\text{ext}}.
\end{equation}
After writing $\Delta^{-1}$ as an integral operator involving the usual
gravitational Green function, the contributions can be analyzed for $r \gg R \geq r'$
through a standard multipole expansion.
The matching to Eq.\ (\ref{PhiEFT}) then implies the identifications
\begin{align}
m = \int d^3 x \, \rho_0 , \quad
Q^{K_l} = \int d^3 x \, \rho_1 r^l \hat{n}^{K_l} ,
\end{align}
as before. We assume that the dipole vanishes, as it cannot be excited
in binary systems; see Eq.\ (\ref{fdipole}). In any case, one can otherwise
redefine the center in the full theory such that the dipole vanishes exactly.
Notice that here the multipoles were defined as formal coefficients
in the effective action, and only the matching related them to integrals
over the mass density of the source.
% The field generated by $\rho_0(r)$ reads
% \begin{equation}
% 4\pi G \Delta^{-1} \rho_0 = - \frac{G}{r} \int d^3 x \, \rho_0 .
% \end{equation}
% \begin{align}
% 4\pi G \Delta^{-1} \rho_1 % &= - \int d^3 x' \, \frac{G \rho_1(\vct{r}')}{|\vct{r} - \vct{r}'|} , \\
% &= - \int d^3 x' \, \frac{G \rho_1(\vct{r}')}{r
%         \sqrt{1 - 2 \vct{n}' \cdot \vct{n} \frac{r'}{r}
%          + \frac{r'^2}{r^2} } } , \\
% \begin{split}
% &= - \frac{G}{2} Q^{ij} \frac{n^i n^j}{r^3}
%         \int d^3 x' \, \rho_1(\vct{r}') r'^2 ( 3 n'^i n'^j - \delta_{ij} ) \nl4
%         + \dots ,
% \end{split}
% \end{align}

The conceptual problem here is that the dynamics of the multipoles is still unknown.
Just the coupling between the multipoles and the gravitational field is fixed for now.
Therefore, we come back to the idea from Sec.\ \ref{int_response} to describe
the dynamical reaction of the multipoles to gravitational interaction
by a response function; see Eq.\ (\ref{eq:quad_sol}) for the quadrupole.
% In fact, it is enough to implicitly specify the
% whole dynamics of the quadrupole. Let us assume for definiteness that
% the dynamics of the quadrupole is given by a
% linear differential operator $\mathcal O^{ij}{}_{kl}$ describing the propagation
% and self interaction of the quadrupole (corresponding to the variation of the
% unknown terms of the action with respect to the quadrupole).
This is almost trivial in the present case, but in general relativity
the definition of source multipoles is fully clarified for test bodies only.
The matching in the effective field theory can potentially generalize this to
the self-gravitating case.

\subsection{Response function}
The generalization of Eqs.\ (\ref{eq:quad_sol}) and (\ref{Enonrel}) to arbitrary multipole order reads
\be\label{multipoleresponse}
\tilde{Q}^{K_l}(\omega) = - \frac{1}{l!} \tilde{F}_l(\omega)
        \mathcal{F}(\hat{\partial}_{K_l} \Phi)(\vct{z}, \omega) ,
\ee
which is suggested by the coupling of the multipoles in the action (\ref{LPP}).
Remember that $\mathcal{F}$ denotes the Fourier transform.
The response function $\tilde{F}_l(\omega)$ is the important ingredient
of the effective theory that we need to obtain from the matching.
Notice that a response function offers a very generic way to encode the
dynamics, which extends far beyond the specific "full" theory considered here.

On the other hand, we have seen that the multipoles in our full theory
are composed as Eq.\ (\ref{composed}),
\begin{align}
\tilde{Q}^{K_l}(\omega) = \sum_n I_{nl} \mathcal{F}\hat{A}_{n K_l} (\omega) .
\end{align}
Now, the amplitudes $\hat{A}_{n K_l}$ satisfy a forced harmonic
oscillator equation (\ref{eq:amplitude_evolution}), in the STF basis and
Fourier domain given by
\begin{equation}
(- \omega^2 + \omega_{nl}^2) \mathcal{F}\hat{A}_{n K_l} = \mathcal{F} \hat{f}_{n K_l}
        = - \frac{I_{n l}}{l!} \mathcal{F}(\hat{\partial}_{K_l} \Phi_{\text{ext}}) ,
\end{equation}
where we used Eq.\ (\ref{eq:def_f}). Combining both equations, we find the
solution for the quadrupole in the frequency domain,
\begin{equation}\label{Qresponsefull}
\tilde Q^{K_l} = - \frac{1}{l!} \left[ \sum_n \frac{I_{nl}^2}{\omega_{nl}^2 - \omega^2} \right]
        \mathcal{F}(\hat{\partial}_{K_l} \Phi_{\text{ext}})(\vct{z}, \omega) .
\end{equation}

Next, we match Eqs.\ (\ref{multipoleresponse}) and (\ref{Qresponsefull}).
This can be done by noting that $\Phi(\vct{z}) = \Phi_{\text{ext}}(\vct{z})$
if the singular self-field is dropped (and similarly for partial derivatives).
The analytic result for the response function finally reads
\begin{equation}\label{Fanalytic}
\tilde{F}_l = \sum_n \frac{I_{nl}^2}{\omega_{nl}^2 - \omega^2} .
\end{equation}
Because of the normalization in Eq.\ (\ref{xinorm}), one must be careful when
analyzing units. One can check that
\be
[ I_{nl} ] = \sqrt{M} L^{l-1} , \quad
[ \tilde{F}_l ] = M L^{2l} ,
\ee
where $M$ is units of mass and $L$ of length.
Remember that we have $c=1$.
We will show that this response function can be directly obtained from
numerical solutions for the exterior gravitational field.
Interestingly, this gives an alternative to determine the
overlap integrals $I_{nl}$, Eq.\ (\ref{Ioverlap}), which appear here
as the coefficients of the poles of the response function.

\section{Numeric Matching\label{numeric}}
Effective field theories not only allow a matching to a known
full theory but also a matching to experimental data. In the
present, section we explore a matching to numerical simulations,
which may be regarded as numerical experiments.
% The current setup is in fact simple enough to allow a reconstruction
% of the full theory from the data.
We argue that the method immediately applies to more complicated
scenarios.

\subsection{Numerical setup}
In this section, we compute the response of a compact configuration to
time-dependent external excitation numerically.
In the Newtonian case, the response should be ``almost trivial,'' in
the sense that the equation for the Newtonian potential is linear
and not time dependent. In contrast to this, the concept of a ``time-dependent''
relativistic response function is much more involved.

% The background quantities are governed by the usual Poisson equation and
% Hydrostatic equilibrium relation:
% \be
% \Delta \psi_0 = 4\pi G \rho_0,\ \frac{1}{\rho}\vec\nabla P  = -\vec \nabla
% \phi_0.
% \ee
% Assuming spherical symmetry and a equation of state of the form $P =
% P(\rho_0)$, these equations are equivalent to the follwing system of second
% order equations:
% \be
% \frac{1}{r^2}\left( r^2\psi_0'\right)'  = 4\pi G\rho_0,\
% \frac{1}{r^2}\left(\frac{r^2  }{\rho } P(\rho_0)\right)' = -4\pi G\rho_0.
% \ee
We first discuss the solution of the background equations (\ref{hydroeq}) and (\ref{backphi}).
For the sake of simplicity, we focus on the particular polytrope
\be
P = k \rho^2 ,
\ee
since in this case, an exact solution is available.
The density profile is given by
\be
\rho_0 (r) = \frac{\rho_c}{K r}\sin K r, \quad
K = \left(\frac{2\pi G}{k}\right)^{\frac{1}{2}} ,
\ee
where $\rho_c$ is the central density.
% , and the mass function reads
% \begin{align}
% M(r) &= 4\pi \int_0^r d r' r'^2 \rho_0(r') \\
%         &= \frac{4\pi \rho_c}{K^3} [ \sin(K r) - K r \cos(K r) ] .
% \end{align}
% Finally the velocity potential is given by $\phi_0=0$ since this is a static
% configuration.

The two perturbation equations encoded by the Lagrangian
from Sec.\ \ref{var_perturbation} read
\begin{align}
\rho_0 \ddot{\vctsym{\xi}}
&= - \rho_0 \nabla \left( - \frac{c_s^2}{\rho_0} \nabla \cdot (\rho_0 \vctsym{\xi})
	+ \Phi_1 + \vct{x} \cdot \ddot{\vct{z}} \right) , \label{xiEOM} \\
\Delta \Phi_1 &= - 4\pi G \nabla \cdot (\rho_0 \vctsym{\xi}) . \label{numfield}
\end{align}
As we consider a single compact object fixed at the coordinate origin,
it holds $\vct{z}=0$, and we dropped the subscript COM. Remember that
the fluid velocity perturbation can be expressed in terms of a potential,
\begin{equation}
\dot{\vctsym{\xi}} = \vct{u}_1 = \nabla \phi_1 . \label{upot}
\end{equation}
Integrating Eq.\ (\ref{xiEOM}) along an arbitrary line, we can conclude that
\begin{equation}\label{numphi}
\dot{\phi}_1 - \frac{c_s^2}{\rho_0} \nabla \cdot (\rho_0 \vctsym{\xi}) + \Phi_1
= \mbox{const}. 
\end{equation}
We can further set the integration constant to zero without loss of generality,
since it can be absorbed into $\phi_1$.
% or $\Phi_1$ are potentials defined up to a constant.

The next step is to separate time, radial, and angular dependence. This
is achieved using Fourier modes and spherical harmonics
$Y^{lm}$,
\begin{align}
 \Phi_1 &= \frac{1}{2\pi} \int d\omega \sum_{lm}e^{i \omega t} \frac{1}{2} h_0^{lm}(r, \omega)
Y^{lm} , \label{h0def} \\
 \phi_1 &= \frac{1}{2\pi} \int d\omega \sum_{lm}e^{i \omega t} \frac{i}{\omega} U^{lm}(r, \omega)
Y^{lm} .
\label{ansatzpert}
\end{align}
The prefactors are chosen to allow an easy comparison to the relativistic
generalization \cite{Chakrabarti:Delsate:Steinhoff:2013:2}.
The subscript on $h_0$ does not denote the perturbation order but the
component of the metric.
We drop the indices $l$, $m$ and the arguments of $h_0$ and $U$ from now on.
% The equation describing the perturbations on a compact object can be found in
% \cite{Rathore:Blandford:Broderick:2004}, but we will use the Newtonian limit
% of the perturbation equations given by \cite{Lindblom:Mendell:Ipser:1997} since it
% allows a more direct connection to the relativist case.
The perturbation equations are then
\begin{gather}
h_0''+\frac{2 h_0'}{r} + h_0 \left[ \frac{4\pi G \rho_0}{c_s^2} - \frac{l(l+1)}{r^2} \right]
        = \frac{8\pi G \rho_0}{c_s^2} U , \label{eq:pertu1} \\
U'' + U' \left[ \frac{2}{r} +\frac{\rho_0'}{\rho_0} \right]
        + U \left[ \frac{\omega^2}{c_s^2} - \frac{l(l+1)}{r^2} \right]
        = \frac{\omega^2}{2 c_s^2} h_0. \label{eq:pertu2}
\end{gather}
Here, Eq.\ (\ref{eq:pertu1}) arises from Eq.\ (\ref{numfield}) with the right-hand side replaced
using Eq.\ (\ref{numphi}), while Eq.\ (\ref{eq:pertu2}) directly derives from Eq.\ (\ref{numphi}).
% where the mass function of the background fluid is defined by
% \begin{equation}
% m(r) = 4\pi \int_0^r d r' r'^2 \rho_0(r') .
% \end{equation}
After choosing boundary conditions at $r=0$, $r=R$, and $r=\infty$,
these equations can be readily integrated numerically.
% % where $c_s^2 = dP/d\rho$ is the speed of sound, $\rho_0$ is the density and $M =
% % \int r^2\sin\theta \rho_0 drd\theta d\varphi$ is the mass function of the
% background fluid.
% The connection between the perturbation functions $\delta U, H_0$ and the more
% conventional variables in Newtonian perturbation theory is given by
% \be
% H_0 = 2 \psi_1\ ,\ \delta U(r) = -i \omega \phi_1,
% \ee
% where $\psi_1$ is the perturbation of the Newton potential and $\phi_1$ is the
% perturbation of the velocity potential.

Regularity at the origin imposes the boundary conditions
\be
h_0 \propto r^l + \mathcal O(r^{l+1}) , \quad
U \propto r^l + \mathcal O(r^{l+1}) ,
\ee
for the perturbation fields. %, where $h_0$ and $u_O$ are real constants.
Regularity at the surface of the star imposes 
\be
2 U(R) - h_0(R) + \frac{2 G m}{R^2 \omega^2} U'(R) = 0 .
\ee
We have given three boundary conditions.
The remaining arbitrariness of the solution is just its overall
normalization, which has no physical significance here
(due to linearity of the perturbation equations).
% we can choose a normalization to set one of the field value, say at the origin. 
It follows that we have enough boundary conditions to uniquely solve the problem
numerically. No further conditions need to be imposed at $r=\infty$.

% The computation is done in \url{math/connection_gr_newton.nb}.

\subsection{Matching the exterior field}
We have derived in Eq.\ (\ref{PhiEFT}) the generic gravitational field
predicted by the effective theory. This field should of course match
the numerically obtained exterior field of the neutron star.
For simplicity, we consider specific values of $l \geq 2$ and $m$, as the
numeric integration decomposes into such sectors. Inserting
Eq.\ (\ref{TaylorPhi}) into Eq.\ (\ref{PhiEFT}), we
then obtain for the field perturbation
\begin{equation}
\Phi_1^{\text{eff}} = \frac{\hat{n}_{K_l}}{l!} \left[- r^{-l-1}  G Q^{K_l} (2l-1)!!
        + r^l (\partial_{K_l} \Phi_{\text{ext}})(\vct{z}) \right] .
\end{equation}
Translated to the function $h_0$ by Eq.\ (\ref{h0def}) and inserting
the definition of the response (\ref{multipoleresponse}), this reads
\begin{equation}\label{h0eff}
h_0^{\text{eff}} = C_{lm} \left[ r^{-l-1}
        \frac{G (2l-1)!!}{l!} \tilde{F}_l(\omega)
        + r^l \right] ,
\end{equation}
where $C_{lm}$ is an overall normalization
factor that can be related to the magnitude of the external
field. The explicit expression reads
$C_{lm} = \frac{2 N_l^2}{l!} \mathcal{Y}^{lm *}_{K_l}
        \mathcal{F}(\hat{\partial}_{K_l} \Phi_{\text{ext}})(\vct{z}, \omega)$,
where Eq.\ (\ref{STFinverse}) was used.
We dropped the summation over $m$, as we focus on a specific value.
Remember that only the external field part
contributes to Eq.\ (\ref{multipoleresponse}), as the other part leads to
singularities, which are dropped.

As the analytic result (\ref{h0eff}) represents a generic vacuum
solution, it is clear that the exterior part of the numeric solution
can be written as
\begin{equation}\label{h0numeric}
h_0^{\text{numeric}} = C \left[ a_l \left( \frac{G m}{r} \right)^{l+1}
        + \left( \frac{r}{G m} \right)^l \right] \quad \text{for } r>R .
\end{equation}
In other words, $a_l$ is proportional to the ratio of the regular and
irregular parts of the potential as $r \rightarrow \infty$.
The part of $h_0$ diverging for large $r$ is coming from the external
gravitational field, whereas the part approaching zero
is due to the multipole of the object. Notice that $a_l$ is dimensionless,
and its definition corresponds to the one in Ref.\ \cite{Damour:Nagar:2009:4} in
the relativistic case, except that here it is a function of $\omega$.
The numeric construction from the last section leads to a unique
numeric value for $a_l$.

Matching the effective potential (\ref{h0eff}) to the exterior numerics
(\ref{h0numeric}) results in
\be
\tilde{F}_l(\omega) = \frac{(Gm)^{2l+1} l!}{G(2l-1)!!} a_l(\omega) .
\ee
Comparing with Ref.\ \cite[Eq.\ (48)]{Damour:Nagar:2009:4} we see that
$\tilde{F}_l = l! \mu_l + \Order(\omega)$, where $\mu_l$ is related to the
dimensionless Love number of the second kind $k_l$ by Ref.\ \cite[Eq.\ (47)]{Damour:Nagar:2009:4}.

\subsection{Numerical results}
We integrated the system of equations \eqref{eq:pertu1} and \eqref{eq:pertu2} with
suitable boundary conditions for a typical mass and radius of a neutron star.
More specifically, we chose parameters such that the radius is $R \approx 8.89$~km and the
mass is $m \approx 1.2 M_\odot$. We considered the quadrupolar case $l=2$ here.

\begin{figure}
\includegraphics{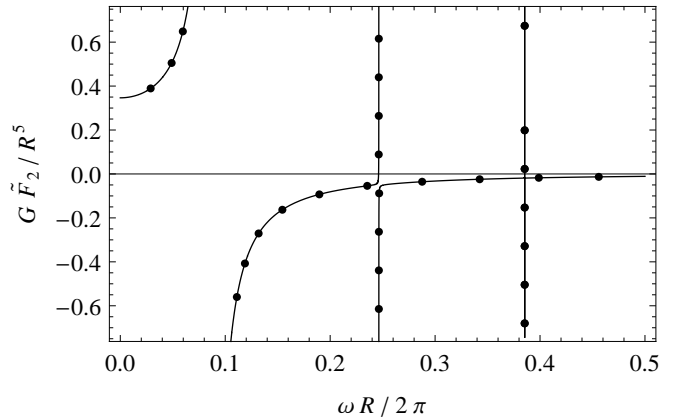}
\caption{Quadrupolar $l=2$ response function for a star with
$R = 8.89$~km, $m = 1.2 M_\odot$, and polytropic index 1 obtained
numerically. The dots are just some selected data points. Many more
were used for the fit, with higher density around the poles.}
\label{fig:GF_newt}
\end{figure}

\begin{table}
\centering
\begin{tabular}{lccc}
% \hline \hline
 Mode   		& $f$	& $p_1$	& $p_2$ \\
 \hline  
%  $\nu_{nl}$~[kHz]	& 2.9464	& 8.3124	&13.0019	\\
%  $I_{nl}^2$~[kg km$^2$]$^2$ & 7.3181+1 & 1.70158-1	&1.58901-3	\\
$\nu_{nl}$~[kHz]	& 2.95	& 8.3	& 13.0	\\
$\omega_{nl} R / 2 \pi$	& 0.0873	& 0.25	& 0.385 \\
$q_{nl}$ & 0.32 & 0.016	& 0.002	\\
% \hline \hline
\end{tabular}
\caption{Frequencies $\nu_{nl} = \omega_{nl} / 2\pi$ and overlap integrals
for a star with $R = 8.89$~km,
$m = 1.2 M_\odot$, and polytropic index 1 obtained from fitting the
response function of the quadrupole, $l=2$.
% $A+B$ stands for $A \times 10^B$.
\label{tab:fit}}
\end{table}

The result is summarized in Fig.\ \ref{fig:GF_newt}, where three poles can
distinctly be seen. Notice that we formed dimensionless quantities using the
size of the object, which is most natural for effective field theories due
to scaling arguments \cite{Goldberger:2007}.
Figure \ref{fig:GF_newt} further suggests that the response function is given by
the superposition of response functions for harmonic oscillators,
\bea
\frac{G \tilde{F}_l}{R^5} = \sum_n \frac{q_{nl}^2}{R^2 (\omega_{nl}^2 - \omega^2)} ,
\label{greenfit}
\eea
which we know must hold exactly from our analytic result (\ref{Fanalytic}).
The dimensionless overlap integrals $q_{nl}$ are related to the $I_{nl}$ through
\begin{equation}
q_{nl}^2 = \frac{G}{R^3} I_{nl}^2 .
\end{equation}
In contrast to this definition, other publications often define dimensionless
overlap integrals based on the central density of the star, while our convention
is adapted to the Love number $k_l$.
We fitted our numerical results using the first three terms of Eq.\ (\ref{greenfit}) and found
the first frequencies of the normal modes ($f$, $p_1$, and $p_2$-modes) and the
associated overlap integrals; see Table \ref{tab:fit}.
Remember that $I_{nl} \geq 0$, and we also assume $q_{nl} \geq 0$.

The numerical matching described here is not only a useful alternative
to obtain mode frequencies and overlap integrals. The method is applicable
in more complicated situations, too, even for full-fledged 3-dimensional
simulations. This is possible as no presuppositions on the response function
are made. Instead, it always comes out as a numeric function, whether one can
find a good and interpretable fit or not.
The exterior potential is always the linear combination
(\ref{h0numeric}) (though sectors with different $l$ do not
decouple in general).

% 
% Comparing to the analytic form of a single mode's Green function, we find that
% the full Green function of the star is given by a superposition of the modes
% individual Green functions. We further directly and simply find the frequencies
% of the modes and the overlap integrals:
%

\section{Conclusions and Outlook\label{conclusions}}
In this paper, we considered astrophysically relevant perturbations of compact
objects in the Newtonian framework. We showed that the effective field
theoretical approach available in the relativistic case applies naturally in
the nonrelativistic case, too. This was expected, since Newtonian
gravitation comes out of general relativity in the appropriate limit. However,
we see here explicitly the connection between an effective description in one
case and an exact rewriting at the level of the action in the other case.

Following the effective field theory approach, we showed how to
describe a generic compact body deformed in a time-dependent way by a
point particle with multipolar degrees of freedom. We argued that the numeric
matching can in principle be applied to arbitrary complicated structured
objects. The method allows a systematic way to understand the impact of
the internal structure on the dynamics of a binary system. This is due to
the fact that the effective theory is matched to a single object first,
which allows one to model a single object (e.g., by mechanical models
like oscillators) as a building block before proceeding to the binary case.
% The interaction in binary systems can then be characterized by potentials
% due to monopole-monopole, monopole-quadrupole (\ref{Vquad}),
% quadrupole-quadrupole, and so forth and so on.
The potentials of a binary system can then be characterized as monopole-monopole, monopole-quadrupole (2.19), quadrupole-quadrupole interactions, and so forth and so on.
This method has interesting analogies to thermodynamics, where systems are
characterized on a macroscopic level by state functions. Indeed, multipoles
encode the macroscopic gravitational interaction of compact objects.
Predictions require the knowledge of correlation functions between state
variables, which is analogous to the response functions here.

The effective description is
explicit once we compare the Newton potential of the actual compact object with
the potential of a multipole alone.
Actually, this matching procedure further provides a prescription to
compute the response of a compact source to an external perturbation. The
perturbation comes as a regular part in the Newtonian potential, while the
backreaction of the central object is irregular at the origin of the
object. From a mathematical perspective, the irregular solution to the Poisson
equation is sourced by a delta distribution located at the center. This is precisely
interpreted as being the potential generated by the source multipole we are
considering. The (properly normalized) ratio of the regular and irregular
contributions is then understood as the response of the object to the external
perturbation. The response function encodes the tidal coefficients of the
central object, which come out as coefficients of a Taylor series in frequency.

Furthermore, our formalism gives a straightforward way to compute the normal
modes of compact objects in Newtonian gravity. Indeed, only regularity
conditions at the origin and at the surface of the star are required. The
generic solution should then be continuously connected to the regular and
irregular solution of the source-free Poisson equation (Laplace equation) in
order to extract the response function. The poles of the response function are
then precisely the normal modes of the compact object, while the overlap
integrals are related to the width of the poles and can be obtained from a
simple fit.

One obvious extension of the present work is to generalize the problem to the
relativistic case. This will be presented in another publication
\cite{Chakrabarti:Delsate:Steinhoff:2013:2} and is based on the numerical
matching method. The results are essentially the
same, in the sense that in the end, the response function turns out to be
related to the ratio of external-field and quadrupole-reaction solutions to the source-free
perturbation equation. However, the problem is much more difficult to attack
since in this case, the equation itself admits singular points, the solution is
expressed as a series of special functions, and the singular effective source has
to be regularized with a suitable renormalization scheme. The Newtonian case is
then very enlightening since it is much easier and not plagued by the same
amount of technical difficulties. For instance, the Hermitian operator $\mathcal{D}$
gives rise to a complete system of modes. No general relativistic analog is
known to us.

Finally, it should be noted that also in the relativistic
case, the fit for the response is to a good approximation a sum of harmonic
oscillators (\ref{greenfit}), at least for the simple neutron star model
considered in Ref.\ \cite{Chakrabarti:Delsate:Steinhoff:2013:2}. This implies that
the quadrupole can be written as a sum of oscillator amplitudes (\ref{composed})
in the relativistic case, with the internal dynamics given by Eq.\ (\ref{LNM}).

\acknowledgments
We gratefully acknowledge fruitful discussions with P.\ Pani and V.\ Cardoso.
We also acknowledge V. Vitagliano for useful expressions.
This work was supported by DFG (Germany) through Projects No.\ STE 2017/1-1 and No.\ STE 2017/2-1,
FCT (Portugal) through Projects No.\ PTDC/CTEAST/098034/2008, No.\ PTDC/FIS/098032/2008,
No.\ SFRH/BI/52132/2013, and No.\ PCOFUND-GA-2009-246542 (cofunded by Marie Curie Actions),
and CERN through Project No.\ CERN/FP/123593/2011.

% \newpage	
\appendix

\section{STF formalism\label{STF}}
In this appendix, we summarize some relations for STF
tensors. Reviews of the STF-tensor formalism and the
relation to spherical harmonics can be found in Sec.\ II of Ref.\ \cite{Thorne:1980},
Appendix A of Re.\ \cite{Blanchet:Damour:1986}, and Sec.\ II of
Ref.\ \cite{Damour:Iyer:1991}.

\subsection{Basic relations}
With the help of a multi-index
$K_l = \{ k_1, k_2, \dots, k_l \}$, we introduce the notation
\begin{align}
n^{K_l} &= n^{k_1} n^{k_2} \dots n^{k_l} , \\
\hat{n}^{K_l} &= [ n^{k_1} n^{k_2} \dots n^{k_l} ]^{\text{STF}} .
\end{align}
The orthogonality property \cite[Eq.\ (2.5)]{Thorne:1980} can be written in the
convenient form
\begin{equation}
\int d \Omega \, \hat{n}^{J_{l'}} \hat{n}^{K_l} = N_l^2 \delta_{l'l}
\hat{\delta}^{J_l}{}_{K_l} , \quad
N_l^2 = \frac{4\pi l!}{(2l+1)!!} , \label{STForth}
\end{equation}
where $\hat{\delta}^{J_l}{}_{K_l}$ is the STF projector, i.e.,
\begin{equation}
[ B^{K_l} ]^{\text{STF}} = \hat{\delta}^{K_l}{}_{J_l} B^{J_l} ,
\end{equation}
for an arbitrary $B^{K_l}$. Using the normalization
\be
\int d\Omega Y^{lm} Y^{l'm'*} = \delta_{mm'}\delta_{ll'},
\ee
the scalar spherical harmonics can be expressed as
\begin{equation}
Y^{lm} = \mathcal{Y}^{lm}_{K_l} \hat{n}^{K_l} , \label{YtoSTF}
\end{equation}
where $\mathcal{Y}^{lm}_{K_l}$ is given by \cite[Eq.\ (2.12)]{Thorne:1980}.
A bijection between the STF-$l$ tensor basis and $m$-basis is provided through
\cite[Eq.\ (2.13)]{Thorne:1980} by virtue of $\mathcal{Y}^{lm}_{K_l}$, e.g., we can
invert Eq.\ (\ref{YtoSTF}) as
\begin{equation}\label{STFinverse}
\hat{n}^{K_l} = \sum_m N_l^2 \mathcal{Y}^{lm *}_{K_l} Y^{lm} .
\end{equation}
From Eq.\ (\ref{STForth}), it follows that
\begin{gather}
N_l^2 \mathcal{Y}^{lm' *}_{K_l} \, \mathcal{Y}^{lm}_{K_l} = \delta_{m'm} , \\
N_l^2 \sum_m \mathcal{Y}^{lm *}_{J_l} \, \mathcal{Y}^{lm}_{K_l} = \hat{\delta}^{J_l}{}_{K_l} .
\end{gather}

We then define the transformation between STF components and $lm$ components by
\begin{equation}
\hat{B}^{K_l} = \sum_m N_l \mathcal{Y}^{lm}_{K_l} B^{lm} , \quad
B^{lm} = N_l \mathcal{Y}^{lm *}_{K_l} \hat{B}^{K_l} .
\end{equation}
Components in the STF basis are often denoted by a hat. Sometimes this notation
is also used for STF projection, but this is always clear from the context.

\subsection{Normal modes in STF basis}
We now transform $\vctsym{\xi}^{\text{NM}}_{nlm}$ and
$A_{nlm}$, reading
\begin{align}
\hat{\vctsym{\xi}}{}^{\text{NM}}_{nl K_l}(\vct{x}) &= \sum_m
N_l \mathcal{Y}^{lm *}_{K_l}
\vctsym{\xi}^{\text{NM}}_{nlm}(\vct{x}) , \\
\hat{A}_{nl K_l}(t) &= \sum_m N_l \mathcal{Y}^{lm}_{K_l} A_{nlm}(t) .
\end{align}
Notice that the complex conjugation in the first equation is due to the fact
that $\vctsym{\xi}^{\text{NM}}_{nlm}$ gives a basis, while $A_{nlm}$ are
components. This allows us to write
\begin{align}
\vctsym{\xi} &= \sum_{nlm} A_{nlm}(t) \vctsym{\xi}^{\text{NM}}_{nlm}(\vct{x}) ,
\\
&= \sum_{nl} \hat{A}_{nl K_l}(t) \hat{\vctsym{\xi}}{}^{\text{NM}}_{nl
K_l}(\vct{x}) ,
\end{align}
Notice that $\hat{\vctsym{\xi}}{}^{\text{NM}}_{nl K_l}$ is real, while
$\vctsym{\xi}^{\text{NM}}_{nlm}$ and $\mathcal{Y}^{lm}_{K_l}$ are complex.
It holds
\begin{equation}
\mathcal{D} \hat{\vctsym{\xi}}{}^{\text{NM}}_{nl K_l} = \omega_{nl}^2
\hat{\vctsym{\xi}}{}^{\text{NM}}_{nl K_l} ,
\end{equation}
where we used that $\mathcal{Y}^{lm}_{K_l}$ provides a bijection between the STF-$l$
tensors basis and $m$-basis. The $\hat{\vctsym{\xi}}{}^{\text{NM}}_{nl K_l}$ are
orthonormal,
\begin{equation}
\int d^3 x \, \rho_0 \hat{\vctsym{\xi}}{}^{\text{NM}}_{n'l' J_{l'}} \cdot
\hat{\vctsym{\xi}}{}^{\text{NM}}_{nl K_l}
	= \delta_{n'n} \delta_{l'l} \hat{\delta}^{J_l}{}_{K_l} .
\end{equation}
This follows from Eq.\ (\ref{xinorm}) and the properties of $\mathcal{Y}^{lm}_{K_l}$
listed above.

\subsection{Useful formulas}
Using above formulas, the angular integration in Eq.\ (\ref{composed})
immediately follows as
\begin{align}
\int d \Omega \, Y_{l'm} \hat{n}^{K_l}
= \mathcal{Y}^{l'm}_{J_{l'}} \int d \Omega \, \hat{n}^{J_{l'}} \hat{n}^{K_l}
= \delta_{l'l} N_l^2 \mathcal{Y}^{lm}_{K_l} . \label{Iangular}
\end{align}
% It is lengthy but easy to check iteratively that
Another useful relation is \cite[Eq.\ (A 34)]{Blanchet:Damour:1986},
\be\label{multipotential}
\partial_{K_l} \frac{1}{r} = (-1)^l \frac{(2l-1)!!}{r^{l+1}} \hat n^{K_l} .
\ee

\section{Overlap in terms of displacement\label{overlapconvention}}
The angular dependence of the displacement vector can be separated as
\begin{align}
\vctsym{\xi}^{\text{NM}}_{nlm} &= \xi^{\text{R}}_{nl}(r) \vct{Y}^{R,lm}(\Omega)
	+ \xi^{\text{E}}_{nl}(r) \vct{Y}^{E,lm}(\Omega) \nonumber \nl
	+ \xi^{\text{B}}_{nl}(r) \vct{Y}^{B,lm}(\Omega) .
\end{align}
The three parts correspond to radial (R), electric-type (E), and magnetic-type
(B) modes with corresponding orthogonal (but un-normalized) vector spherical harmonics,
\begin{align}
\vct{Y}^{E,lm}(\Omega) &= r \nabla Y^{lm}(\Omega) , \\
\vct{Y}^{B,lm}(\Omega) &= \vct{n} \times \vct{Y}^{E,lm}(\Omega) , \\
\vct{Y}^{R,lm}(\Omega) &= \vct{n} Y^{lm}(\Omega) .
\end{align}
% see (2.18) in \cite{Thorne:1980}.
Notice that the radial functions are independent of $m$.

As $\vctsym{\xi}$ can be related to a scalar potential (\ref{upot}),
we must have $\xi^{\text{B}}_{nl} = 0$. From Eqs.\ (\ref{rhodecomp}) and
(\ref{rhoseparate}), we then obtain
\begin{equation}
\rho^{\text{NM}}_{nl} = - \frac{1}{r^2} \frac{d ( r^2 \rho_0 \xi^{\text{R}}_{nl} )}{d r}
        + \frac{1}{r} \rho_0 l (l+1) \xi^{\text{E}}_{nl} .
\end{equation}
Our convention for the overlap integrals (\ref{Ioverlap}) now read
\begin{equation}
I_{nl} = N_l l \int d r \, r^{l+1} \rho_0
        \left[ \xi^{\text{R}}_{nl} + (l+1) \xi^{\text{E}}_{nl}\right] .
\end{equation}

\ifnotprd
\bibliographystyle{utphys}
\fi

\ifarxiv
% run: cp NSresponseNewton.bbl NSresponseNewton_refs_arxiv.tex
\input{NSresponseNewton_refs_arxiv}
\else
\bibliography{QNM_resonance_PN}
\fi

\end{document}

%% file: NSresponseNewton_refs_arxiv.tex
\providecommand{\href}[2]{#2}\begingroup\raggedright\endgroup